\documentclass[sn-mathphys-ay]{sn-jnl}

\usepackage{graphicx}%
\usepackage{multirow}%
\usepackage{amsmath,amssymb,amsfonts}%
\usepackage{amsthm}%
\usepackage{mathrsfs}%
\usepackage[title]{appendix}%
\usepackage{xcolor}%
\usepackage{textcomp}%
\usepackage{manyfoot}%
\usepackage{booktabs}%
\usepackage{algorithm}%
\usepackage{algorithmicx}%
\usepackage{algpseudocode}%
\usepackage{listings}%
\usepackage{natbib}

\theoremstyle{thmstyleone}%
\newtheorem{theorem}{Theorem}
\theoremstyle{thmstyletwo}%

\theoremstyle{thmstylethree}%

\newtheorem{axO}{Axiom TO}
\newtheorem{axN}{Axiom TN}
\newtheorem{prop}{Proposition}
\newtheorem{lemma}{Lemma}

\begin{document}

\title[Physical dimensions]{Species of structure and physical dimensions}


\author[]{\fnm{Heinz-J\"urgen} \sur{Schmidt}}

\affil[1]{\orgdiv{Fachbereich Mathematik/Informatik/Physik}, \orgname{Institut f\"ur Physik}, \orgaddress{\street{Universit\"at Osnabr\"uck}, \city{Barbarastr. 7}, \postcode{D-49069}, \state{Osnabr\"uck}, \country{Germany}}}


\abstract{This study addresses the often underestimated importance of physical dimensions
and units in the formal reconstruction of physical theories, focusing on
structuralist approaches that use the concept of ``species of structure" as a meta-mathematical tool.
Similar approaches also play a role in current philosophical debates
on the metaphysical status of physical quantities.
Our approach builds on an earlier proposal by Terence Tao.
It involves the representation of fundamental physical quantities by one-dimensional
real ordered vector spaces, while derived quantities are formulated using
concepts from linear algebra, e.~g.~tensor products and dual spaces.
As an introduction, the theory of Ohm's law is considered.
We then formulate a reconstruction of the calculus of physical dimensions,
including  Buckingham's $\Pi$-theorem.
Furthermore, an application of this method to the Newtonian theory of
gravitating systems consisting of point particles is demonstrated,
emphasizing the role of the automorphism group and its physical interpretations.}

\maketitle

\section{Introduction}\label{sec:Intro}

Physical quantities are not simply numbers, but have a physical dimension
and can therefore be represented by products of numbers with physical units.
This is a fundamental fact that is already taught to pupils in physics lessons.
It is therefore surprising that this topic hardly plays a role in the mathematical
formulation of physical theories.  Obviously, you can calculate with physical quantities,
they can be added, multiplied and divided to form new quantities.
This raises the question of the mathematical structure of these operations,
which we will summarize here as ``dimensional calculus".
There are numerous approaches to this, some of which have led to elaborate theories of dimensional calculus,
of which we would like to mention \citet{W68},  \citet{R18} and \citet{D24}.
In addition, the ideas from the weblog of Terence Tao, see \citet{T13}, should be mentioned here,
who does not define any new algebraic structures for the dimensional calculus,
but relies on well-known structures of linear algebra such as tensor product and dual space.
In our work, among other things, a dimensional calculus is sketched that incorporates
these ideas of \citet{T13} and develops them further in some places,
for example concerning the extension to rational exponents of dimensional quantities
and the treatment of the $\Pi$-theorem.

However, one can ask the question whether the reconstruction of the structure of physical quantities
in general physical theories is a task of mathematical physics at all, and not rather a task of formal  philosophy of science.
In the following, we will limit ourselves to approaches to the reconstruction of
physical theories that can be described as ``structuralistic" in a broad sense, see  \citep{S24}.
In the corresponding main works we find on the one hand indications
that in the reconstructions of physical theories ``space" and ``time"
cannot simply be modeled by ${\mathbb R}^3$ and
${\mathbb R}$, see \citep[p.~88]{L90}  and  \citep[p.~30 ff.]{BMS87}.
On the other hand, the special character of dimensional quantities
is not consistently taken into account, for example when introducing
masses as positive real numbers, see  \citep[p.~90]{L90}, or \citep[ p.~31]{BMS87}.

What the structuralistic approaches considered in  \citep{S24} have in common is
that they use the meta-mathematical tool of so-called
``species  of structure" for the reconstruction of physical theories.
The concept of ``species  of structure" is initially based on the insight,
which goes back to N. Bourbaki, that mathematical theories such as "Lie groups" or "Linear algebra"
can be understood as theories of certain species  of structure.
This concept develops its real effectiveness when it is extended to
relationships between species  of structure, which will, however,
not play a role in this paper. In the context of physics,
it is now crucial to realize that physics not only makes use of mathematics in a general sense,
but that to every physical theory $T$ can be assigned a species  of structure $\Sigma$.
Among other things, $\Sigma$ contains a collection of base sets,
some of which can be given a physical meaning.

Therefore, in this paper we take the position that physical quantities
must be represented by base sets of $\Sigma$ and suitable structural elements and axioms
and make suggestions for this. Ranges of physical quantities are represented
by one-dimensional ordered real vector spaces and the calculus of physical quantities
will be reconstructed by employing the mathematical concepts from linear algebra
of tensor products of vector spaces and the dual of a vector space
as suggested by \citet{T13}.
The embedding of the dimensional calculus in the meta-theory of the
species of structure does not serve its more transparent formulation
but is a necessary extension of this meta-theory for the case of application to physical theories.
However, we cannot present a complete theory of physical dimensions here,
but must limit ourselves to the most important cases and examples.

In view of the fact that the topic of dimensional quantities is largely neglected in mathematical physics,
it is all the more remarkable that this very topic plays a role in some recent
philosophical debates on the metaphysical status of physical quantities, see,
for example, the book by  \citep{W20} entitled ``The Metaphysics of Quantities".
Chapter 10 of this book deals with ``a structuralist understanding of quantities",
although the term "structuralism" used there refers rather to metaphysical positions
such as ``structural realism" and is not to be equated with the structuralistic tradition outlined in \citet{S24}.

One branch of this debate concerns the status of concepts such as ``mass" and moves along,
analogous to earlier debates about the status of space and time,  a spectrum of positions
between ``absolutism" and ``comparativism".
To illustrate the influence and scope of this debate,
the following works can be cited as examples: \citet{S09},  \citet{L12},
 \citet{D13}, \citet{B20},  \citet{M21}, \citet{M22},
 \citet{MJ23}, \citet{CJ23}, \citet{D24}, and \citet{M24}.

The present paper is not intended to formulate and defend positions on this debate.
Instead, we recommend using in this debate the tools of the partially forgotten structuralist tradition
and its further development outlined here.
We would like to illustrate this recommendation with an important example.
In the debate about the metaphysical status of physical concepts
the term "symmetry" plays a key role. In the historical debate about absolute vs.~relative space
Leibniz had put forward the argument that a constant displacement of all bodies relative
to absolute space should make no difference. Similarly, in the current debate,
the effect of a constant dilation of all masses has been discussed.
What role does it play that the escape velocity of a body on the earth's surface
is not invariant under mass dilations, see \citet{B20}? Or does the gravitational constant
has to be transformed as well according to  \citet{CJ23}, which in turn leaves the escape velocity unchanged?
My impression of this debate is that it could benefit from the clarification
of the concept of symmetry that is possible using the structuralist reconstruction
of physical theories, see Section \ref{sec:AP}.
It has been noted in \citep[section 6.5]{M24} that
\begin{quote}
 Whether absolute masses are or are not dynamically relevant,
 which is the main point of contention between realists and anti-realists,
 depends on the laws of nature: their syntactical form (i.~e.~the equations),
 how one interprets that form, and how one interprets constants of nature that feature in the laws (...).
\end{quote}
In my opinion, this statement illustrates the relevance for the current debate of a formal reconstruction of
Newton's theory of gravitation including the fundamental quantities, as attempted in the present work.

An obvious advantage of metatheoretic frameworks is that they make questions mathematically decidable
that were previously controversial, such as the question of the uniqueness of the gravitational constant,
see Section \ref{sec:NGPP}.
However, the choice of a particular framework can also be controversial.
One cannot rule out the possibility that a formal framework for reconstructing theories
makes implicit metaphysical assumptions and is therefore no longer ``unbiased" when used to deciding on such issues.
This is particularly likely in the case of {\it structuralist} frameworks,
which may tend towards the comparativist side in the absolutism/comparativism debate.
But even that would be an interesting result, because the metaphysical status of
physical quantities is not  {\it ad hoc} built into the structuralist concept.

The paper is organized as follows. In Section \ref{sec:Ohm} we consider a physical theory $TO$
designated to formulate Ohm's law.
As simple as this theory is, it already illustrates a fundamental issue
in the theory of physical quantities,
namely how a new quantity with the derived dimension ``resistance"
can be obtained from two fundamental quantities such as
``voltage" and ``current" on the basis of a law, in this case Ohm's law.

This approach is used in Section \ref{sec:CPQ} to sketch a general dimensional calculus
based on the ideas of \citet{T13}, using structures from linear algebra.
The well-known $\Pi$-theorem is reformulated in this framework, see subsection \ref{sec:BPT},
and proved in the appendix \ref{sec:Pi}.
To do this, we use the affine geometry of positive ranges of quantities
given by the exponential action of real numbers.
The $\Pi$-theorem describes how a physically meaningful equation involving
dimensional quantities can be rewritten as an equation involving only dimensionless quantities.
Also $3$-dimensional vector quantities are considered in Section \ref{sec:VQ}
since they are required for Section \ref{sec:NGPP}.
The next Section \ref{sec:SS} contains an informal account of the notion of ``species of structure" and
an example of how it would be defined for Ohm's theory of Section \ref{sec:Ohm}.
After a general description in subsection \ref{sec:SSGA} we describe, in subsection \ref{sec:SSPT},
the special case where a physical law in a species-of-structure theory
falls within the scope of the $\Pi$-theorem.
In Section \ref{sec:NGPP} we outline a ``real" physical theory,
namely Newton's theory $TN$ of $N$ gravitationally interacting point particles
using the formalism developed in Section \ref{sec:CPQ}.
There is some similarity between Ohm's law $U=R\,I$ and the equation of motion in $TN$,
saying that the force on the $i$th particle is the product of $\Gamma$ times
the sum of the contributions due to attractions from the other particles.
But in this case the gravitational constant $\Gamma$ is {\em universal} and could, in principle,
be used to reduce the number of fundamental quantities.
After the general account of $TN$ in subsection \ref{sec:NGGA} we more closely
investigate isomorphisms and automorphisms in the corresponding species-of-structure theory
generated by dilations of the fundamental quantities ``length", ``time" and ``mass",
see subsection  \ref{sec:NGIA}.

In section \ref{sec:AP} we outline three physical interpretations
of the notion of an ``automorphism" of a structure of the species $\Sigma$,
referred to as ``Leibniz interpretation", ``active'' and ``passive interpretation",
which could probably be used to clarify the concept of symmetry that has emerged
in recent debates in the philosophy of science.
We close with a summary and outlook in Section \ref{sec:SO}.
In Appendix \ref{sec:ACP}
we recapitulate an axiomatic characterization of positive domains and their embedding
into signed ranges of physical quantities.
The Appendix \ref{sec:RE} contains a proposal for a definition of rational
powers of physical quantities, which, however, is not needed for the rest of the paper.

\section{Ohm's law $U=R\,I$}\label{sec:Ohm}

We consider a highly simplified physical toy theory $TO$,
which is essentially built around Ohm's law.
This theory will be used to explain the basic definitions
for "ranges of physical quantities".

\begin{figure}[ht!]
\centering
\includegraphics*[clip,width=0.7\columnwidth]{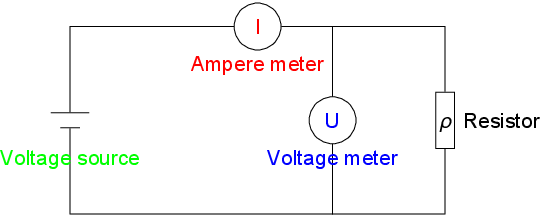}
\caption{Schematic sketch of an electrical circuit for carrying out measurements to confirm Ohm's law.
}
\label{FIGOHM}
\end{figure}

We consider a set of electrical resistors that are
produced using some established methods,
which is represented by a base set ${\sf R}$.
Various voltages are applied to these resistors, which are measured
together with the resulting electrical currents, see Figure \ref{FIGOHM}.
We assume that the values obtained by these measurements are elements
of two other basis sets ${\mathcal U}$ for voltages and ${\mathcal I}$ for currents,
which have the structure of one-dimensional, real, ordered vector spaces.
Their order can be defined by distinguishing the rays ${\mathcal U}_{>0}$,
resp.~${\mathcal I}_{>0}$, which consist of all positive elements.
Physical units such as ``Volt" or ``Amp\`ere" are positive bases in ${\mathcal U}$  or, resp.~, ${\mathcal I}$.
They consist of a singleton since the spaces are one-dimensional.
The value of a physical quantity like ``voltage" can hence be written in the familiar way
as a positive number times a physical unit.
The question of how to construct these variables is not addressed in this paper.
This would be the subject of measurement theories.
We emphasize that the set ${\sf R}$ of resistors is not understood as a range of quantities,
mathematically it is just an unstructured set. Therefore, we denote the elements of
${\sf R}$ with Greek letters as $\rho$ and not with $R$, which could mean the value of a quantity with dimension.

The set of all possible measurements
will be mathematically represented by a subset
\begin{equation}\label{Gamma}
  \gamma \subset {\sf R}\times{\mathcal U}_{>0}\times {\mathcal I}_{>0}
  \;.
\end{equation}

It turns out to be advantageous to abstract from the differences between different resistors
in the example theory $TO$, insofar as they do not show up in the experiments outlined above.
The elements of ${\sf R}$ are therefore not to be interpreted as concrete resistors,
but as certain equivalence classes of these.
Therefore, we require that two resistors are equal if they have the same ``characteristic curve":
\begin{axO} \label{Axiom1}
For all $\rho_1,\rho_2\in{\sf R}$ there holds:
\begin{equation}\label{A11}
 \{(U,I)\left| \left(\rho_1,U,I \right)\in \gamma\right.\})=
 \{(U,I)\left| \left(\rho_2,U,I \right)\in \gamma\right.\}
 \Rightarrow\rho_1=\rho_2
 \;,
\end{equation}
\end{axO}

We now turn to the problem of how to formulate the experimental finding
that $U$ is proportional to $I$ in different experiments with the same resistor $\rho$.
To this end we consider linear mappings $m:{\mathcal I} \rightarrow {\mathcal U}$.
The space of these mappings is again one-dimensional and can be identified
with ${\mathcal U} \otimes {\mathcal I}^\ast$, where $\otimes$ denotes the
tensor product of vector spaces and $ {\mathcal I}^\ast$ the dual vector space
defined as the space of linear mappings $\imath: {\mathcal I}\rightarrow {\mathbb R}$.
In linear algebra this is the usual identification of second rank tensors possessing
one covariant index and one contravariant index with the matrix of a linear map.
Because of the one-dimensionality,
the elements of ${\mathcal U} \otimes {\mathcal I}^\ast$ can be written in the form
$U\otimes \imath$, with $U\in{\mathcal U}$ and $\imath\in{\mathcal I}^\ast$.
This representation is only unique up to the scale transformation $U\mapsto \alpha U$
and $\imath \mapsto \frac{1}{\alpha}\imath$ for arbitrary $\alpha\in{\mathbb R},\,\alpha\neq 0$.
A linear map $m:{\mathcal I} \rightarrow {\mathcal U}$ is called ``monotone" iff
it maps the positive range ${\mathcal I}_{>0}$ onto ${\mathcal U}_{>0}$.

With this preparation we will formulate the main axiom of $TO$:\\

\begin{axO} \label{Axiom2}
\begin{enumerate}
  \item $\gamma$ is not empty and for all $\rho\in{\sf R}$ there exists a linear, monotone map $m:  {\mathcal I} \to  {\mathcal U}$
    such that
    \begin{equation}\label{axiom2}
        \left(\rho,U,I \right) \in \gamma \Leftrightarrow U= m(I)
        \;.
    \end{equation}
  \item Conversely, for every linear, monotone map $m:  {\mathcal I} \to  {\mathcal U}$ there exists
     a $\rho\in{\sf R}$ such that (\ref{axiom2}) holds.
\end{enumerate}

\end{axO}

This axiom says that, first, for a fixed resistor $\rho$, the measured values of $(U_1,I_1)$ lie on a
line passing through the origin. Second, the slope $\frac{\Delta U}{\Delta I}$ of the line has the value
$U_1\otimes\imath_1\in {\mathcal U} \otimes {\mathcal I}^\ast$, where $\imath_1 \in {\mathcal I}^\ast$
is defined by $\imath_1(I_1)=1$.
Upon the identification of the space $\mbox{Lin}({\mathcal I}\to{\mathcal U})$ of linear mappings
with ${\mathcal U}\otimes {\mathcal I}^\ast$ we can set $U_1\otimes\imath_1=m$.
This slope $m$ is not a real number
but an element of ${\mathcal R}:={\mathcal U} \otimes {\mathcal I}^\ast$, which is an ordered, linear one-dimensional
vector space that can be viewed as a derived range of a new physical quantity.
This quantity is, of course, the electrical resistance.
Since $\gamma\neq \emptyset$ the slope of the line corresponding to the resistor $\rho\in{\sf R}$ is unique
and Axiom $TO$ \ref{Axiom2} enables us to
define a map
\begin{equation}\label{resmp}
{\sf  m}: {\sf R}\rightarrow {\mathcal R}_{>0},\quad \rho\mapsto m \mbox{ satisfying } (\ref{axiom2})
 \;,
\end{equation}
that assigns a resistance to each resistor.
It will be bijective due to Axiom $TO$ \ref{Axiom2}, 2nd part.
Nevertheless, it will be useful not to identify ${\sf R}$ and ${\mathcal R}_{>0}$
in order to be able to make certain distinctions between isomorphisms and automorphisms
of the corresponding species of structure, see Section \ref{sec:SS}.

Introducing positive basis vectors, say, ${\rm Volt}\in {\mathcal U }$ and  $\mbox{Amp\`ere}\in {\mathcal I}$
we obtain, first,  a dual basis $\mbox{Amp\`ere}^{-1}\in {\mathcal I}^\ast$ defined by
$\mbox{Amp\`ere}^{-1}(\mbox{Amp\`ere})=1$ and, second, a product basis
${\rm Ohm}:={\rm Volt}\otimes\mbox{Amp\`ere}^{-1}\in {\mathcal U }\otimes{\mathcal I}^\ast$.

The formalism of dual spaces and tensor product spaces of one-dimensional real vector spaces
thus allows a reconstruction of the familiar calculus of multiplication and division of physical units.
This is discussed in more detail in the next section.

\section{Calculus of physical quantities and Buckingham's $\Pi$-theorem}\label{sec:CPQ}

\subsection{Scalar quantities}\label{sec:SQ}

Generalizing the situation of the theory $TO$ sketched in the preceding section,
we consider the case where a physical theory contains $N$  ranges of fundamental scalar physical
quantities ${\mathcal R}_i\,\; i=1,\ldots,N$.
These are assumed to be sets of values of the corresponding quantity
equipped with the structure of one-dimensional ordered, real vector spaces.
The order is given by subsets ${\mathcal R}_{i,>0}$ consisting of the positive values of the
corresponding quantity, as in the previous Section.
In this respect the present approach, while following
\citet{T13}, differs from other reconstructions  considering only positive ranges without
embedding them into vector spaces, as \citet{MJ23}, \citet{CJ23} and  \citet{D24}.
Our approach is particularly suitable for the reconstruction of
Newton's theory of $N$ gravitating particles, see Section \ref{sec:NGPP}.

In section \ref{sec:Ohm} we assumed the basic quantities
``voltage" and ``current" without addressing the question of how
their structure can be physically derived.
This problem is not dealt with in this paper so as not to overload the presentation.
A general scheme for such derivations is known as the
``representational theory of measurement" \citet{M87},
see also the theory of measurement presented in the three volumes of  \citep{KLST71},
\citep[appendix A]{F07} and  \citep[section 2.4]{B15}.

From these fundamental quantities one can construct derived quantities just as the
quantity ``resistance" from the  fundamental quantities ``voltage" and ``current" in the
preceding Section. For this construction we will use repeatedly the tensor product
${\mathcal R}_i\otimes{\mathcal R}_j$ and the dual space ${\mathcal R}_i^\ast$.
${\mathcal R}_i^\ast$ is also a one-dimensional ordered, real vector space if we define
\begin{equation}\label{defdualorient}
 \mbox{For all}\;\rho\in {\mathcal R}_i^\ast :\; \rho>0 \quad
 \stackrel{\rm def}{\Leftrightarrow}\quad \rho(R)>0 \; \mbox{for all}\; R\in R_{i,>0}
  \;.
\end{equation}

The difference to the dimensional calculus presented in \citet{D24} is that in this
work, e.~g., the derived quantity resulting from the multiplication of two quantities
is constructed as a quotient of the Cartesian product of two positive ranges.
This construction exactly mimics the tensor product of two one-dimensional spaces,
as it must be if both approaches are valid.
The same applies to the quotient of quantities and the dual space.

In the general case we have to take account of two complications. First, we will not distinguish between the
two ranges of quantities ${\mathcal R}_i\otimes{\mathcal R}_j$ and ${\mathcal R}_j\otimes{\mathcal R}_i$,
see also the corresponding discussion in \citet{T13}.
More generally, $k$-fold tensor products of ranges of fundamental quantities have to be
distinguished only modulo permutations.
This can be achieved by selecting a fixed sequence of base sets.
Second, physical units can be cancelled. As far as I can see, this problem has not be considered in \citet{T13}.
This means for our formalism that the one-dimensional real vector space
${\mathcal R}_i\otimes{\mathcal R}_i^\ast$ has to be identified with ${\mathbb R}$
by means of the ``canonical" isomorphism ${\sf C}:{\mathcal R}_i\otimes{\mathcal R}_i^\ast\rightarrow {\mathbb R}$
(${\sf C}$ for cancellation)
defined by
\begin{equation}\label{defiso}
 {\sf C}(R\otimes \rho ):= \rho(R)
 \;,
\end{equation}
where $R\in {\mathcal R}_i,\,\rho\in {\mathcal R}_i^\ast$ and the
definition (\ref{defiso}) does not depend on the
chosen tensor factors, i.~e., ${\sf C}\left(\alpha R \otimes \frac{1}{\alpha}\rho\right)={\sf C}(R\otimes \rho ),\,\alpha>0$.
This canonical isomorphism can be easily extended to multiple tensor products of ${\mathcal R}_i$ and ${\mathcal R}_i^\ast$.
To simplify the notation we will write
\begin{eqnarray}\label{Riell1}
 {\mathcal R}_i^{\ell}&:=&\underbrace{{\mathcal R}_i\otimes\ldots \otimes {\mathcal R}_i}_{\ell\, {\rm factors}}\;,\\
 \label{Riell2}
 {\mathcal R}_{i,>0}^{\ell}&:=&\underbrace{{\mathcal R}_{i,>0}\otimes\ldots \otimes {\mathcal R}_{i,>0}}_{\ell\, {\rm factors}}
 \;,
\end{eqnarray}
and
\begin{eqnarray}\label{Rik1}
 {\mathcal R}_i^{-k}&:=&\underbrace{{\mathcal R}_i^\ast\otimes\ldots \otimes {\mathcal R}_i^\ast}_{k\, {\rm factors}}\;,\\
 \label{Rik2}
 {\mathcal R}_{i,>0}^{-k}&:=&\underbrace{{\mathcal R}_{i,>0}^\ast\otimes\ldots \otimes {\mathcal R}_{i,>0}^\ast}_{k\, {\rm factors}}
 \;.
\end{eqnarray}
Then the extension of ${\sf C}$ can be written compactly as
\begin{equation}\label{defisoc}
 {\sf C}: {\mathcal R}_i^\ell \otimes  {\mathcal R}_i^{-k} \rightarrow {\mathcal R}_i^{\ell-k}
 \;,
\end{equation}
for arbitrary natural numbers $\ell,k$ and setting ${\mathcal R}_i^{0}:={\mathbb R}$.
The further extension to the case of integers $\ell,k$ is obvious.

According to the preceding remarks and definitions we define a ``derived quantity"
with values $Q^{({\mathbf n})}$ being elements of the one-dimensional real space
\begin{equation}\label{defderq}
  {\mathcal R}^{({\mathbf n})}:= \bigotimes_{i=1}^N {\mathcal R}_i^{n(i)}
  \;,
\end{equation}
where ${\mathbf n}=( n(1), \ldots,n(N))$ is an $N$-tuple of integers, i.~e., ${\mathbf n}\in {\mathbb Z}^N$.
Thus the range $ {\mathcal R}^{({\mathbf n})}$ of a derived quantity is completely characterized
by the $N$-tuple ${\mathbf n}$. Fundamental quantities with range $R_j$ are special cases of
derived quantities obtained by setting $n(i)=\delta_{ij}$.

The order of ${\mathcal R}^{({\mathbf n})}$ is given by the subset of positive quantities
\begin{equation}\label{deforientderq}
  {\mathcal R}^{({\mathbf n})}_{>0}:=\bigotimes_{i=1}^N {\mathcal R}_{i,>0}^{n(i)}
  \;.
\end{equation}

Next we define the product of the values of two derived quantities $Q^{({\mathbf n})}, Q^{({\mathbf m})}$.
\begin{equation}\label{product}
 Q^{({\mathbf n})}\, Q^{({\mathbf m})} := \bigotimes_{i=1}^N {\sf C}\left(R_i^{n(i)}\otimes R_i^{m(i)} \right)
 \in \bigotimes_{i=1}^N  {\mathcal R}_i^{(n+m)(i)} =  {\mathcal R}^{({\mathbf n}+{\mathbf m})}
 \;.
\end{equation}
In special cases the product of quantities is simply the tensor product, but in general the
cancellation operator ${\sf C}$ is additionally involved.

The inverse of a derived quantity will be defined in three steps. First, we define
the inverse  $R_i^{-1}$ of the value of a fundamental quantity $R_i\neq 0$ by the unique element of the dual space
${\mathcal R}_i^\ast$ satisfying
\begin{equation}\label{definvfund}
 R_i^{-1} (R_i)=1
 \;.
\end{equation}
Second, this definition will be extended to tensor products $R_i^{n(i)},\; n(i)>0$ of values of fundamental quantities  by
\begin{equation}\label{definvprod}
  \left(R_i^{n(i)} \right)^{-1}:=R_i^{-n(i)} := \underbrace{R_i^{-1}\otimes\ldots\otimes R_i^{-1}}_{n(i)\; \mbox{factors}}
  \;,
\end{equation}
and further be extended to tensor products $R_i^{n(i)},\; n(i)<0$ by interchanging the
role of a fundamental range and its dual.
For  $n(i)=0$, i.~e., for a dimensionless quantity the inverse is just the usual inverse of real numbers.
Finally, the inverse of the value of a derived quantity $Q^{({\mathbf n})}$
is defined by
\begin{equation}\label{definvder}
 \left(  Q^{({\mathbf n})}\right)^{-1}:=  Q^{(-{\mathbf n})}:= \bigotimes_{i=1}^N R_i^{-n(i)}
 \in {\mathcal R}^{(-{\mathbf n})}
 \;.
\end{equation}

By combining the two defined operations, multiplication and inversion, we may form arbitrary
``rational monomials" of a finite number of derived quantities consisting of products
of positive and negative powers of their values. Recall that, after introducing physical units,
each value of a derived quantity $Q^{({\mathbf n})}$
can be written as the product of a real number $x^{({\mathbf n})}\in {\mathbb R}$
and a base vector (physical unit)  $U^{({\mathbf n})}$ in
the one-dimensional vector space ${\mathcal R}^{({\mathbf n})}$. The multiplication and inversion
of physical units can be mapped onto addition and reflection
(i.~e.~${\mathbf n}\mapsto -{\mathbf n}$) in ${\mathbb Z}^N$. Further the real numbers
$x^{({\mathbf n})}\in {\mathbb R}$ are multiplied and inverted in the usual way.
The algebraic structure of physical quantities that can be derived from $N$ fundamental
quantities generated by the two operations multiplication and inversion is thus isomorphic to
the algebraic structure of ${\mathbb R}\times {\mathbb Z}^N$.
It is similar to the structure of the ${\mathbb Z}$-module ${\mathbb Z}^N$,
but modified by the restriction of the inversion due to the condition $x^{({\mathbf n})}\neq 0$.

For later purposes we mention the square root of physical quantities.
Let  ${\mathcal R}^{({\mathbf n})}$ be the range of a certain quantity, then the map
\begin{equation}\label{defsqr}
  \sqrt{}: {\mathcal R}^{({\mathbf n})}_{>0}\otimes{\mathcal R}^{({\mathbf n})}_{>0} \rightarrow {\mathcal R}^{({\mathbf n})}_{>0}
\end{equation}
can be defined as follows. Let $Q_1\otimes Q_2$ be an arbitrary element of
${\mathcal R}^{({\mathbf n})}_{>0}\otimes {\mathcal R}^{({\mathbf n})}_{>0}$, i.~e.,
$Q_1,Q_2\in {\mathcal R}^{({\mathbf n})}_{>0}$.
Since  ${\mathcal R}^{({\mathbf n})}$ is one-dimensional, there exists a real number $\lambda>0$ such that
$Q_2=\lambda\,Q_1$. Then one defines
\begin{equation}\label{defsqrt}
 \sqrt{Q_1\otimes Q_2}=\sqrt{\lambda\,Q_1\otimes Q_1}:=\sqrt{\lambda}\,Q_1\in {\mathcal Q}^{({\mathbf n})}_{>0}
 \;.
\end{equation}

\subsection{Buckingham's $\Pi$-theorem}\label{sec:BPT}

We return to the general case. Although Buckingham's $\Pi$-theorem is formulated
and proven at many places, see, e.~g., \citet{G11}, it may be worthwhile to reconsider
it using the present framework. Most published versions of the $\Pi$-theorem use
the parametric form of dimensional calculus; here we will present the $\Pi$-theorem
using the abstract form, which looks similar despite a different meaning of the expressions.
A simple example will be given below.

Assume that we have $N$ fundamental quantities with ranges ${\mathcal R}_i,\,i=1,\ldots,N$, and additionally
define $L$  derived quantities with values $Q_\ell\in {\mathcal Q}_\ell$  of the form
\begin{equation}\label{Qell}
 Q_\ell =\prod_{i=1}^N R_i^{A_{i \ell}},\quad \ell=1,\ldots,L \mbox{ and } A_{i \ell}\in{\mathbb Z}
 \;.
\end{equation}
We use ${\mathcal Q}_\ell$ for the ranges of the derived quantities instead of
${\mathcal R}^{({\mathbf m})}$ as in the last subsection.
Note that (\ref{Qell}) is not a product of real numbers, as it would appear in the parametric form of the $\Pi$-theorem,
but a product of positive or negative powers of fundamental quantities as a special case of
the product of derived quantities defined in the preceding subsection.
The integers $A_{i \ell}$ can be viewed as entries of an $N\times L$-matrix $A$
that can be identified with a linear map $A: {\mathbb R}^L\to{\mathbb R}^N$.
Let $K$ denote the dimension of the kernel (or null space) of $A$. The transposed matrix
$A^\top$ can be identified with a linear map $A^\top: {\mathbb R}^N\to{\mathbb R}^L$.

The (Abelian, multiplicative) group of dilations
${\sf D}:= \{{\boldsymbol\lambda}=(\lambda_1,\ldots, \lambda_N)\left|\right. \lambda_i>0 \}$
operates on the Cartesian product ${\mathcal R}_+:=\prod_{i=1}^N {\mathcal R}_{i,>0}$ in the obvious way,
by the multiplication $R_i \mapsto \lambda_i\,R_i$ in each factor of the product.
This induces a corresponding action of ${\sf D}$ on the derived quantities via
\begin{equation}\label{actionQ}
  Q_\ell \mapsto \Lambda_\ell\, Q_\ell,\quad \mbox{where}\quad \Lambda_\ell = \prod_{i=1}^N \lambda_i^{A_{i\ell}},
  \; \ell=1,\ldots,L
  \;.
\end{equation}
This induced action can be applied to arbitrary elements of the Cartesian product
\begin{equation}
{\mathcal Q}_+:=\prod_{i=1}^L {\mathcal Q}_{\ell,>0}
\;.
\end{equation}
A subset $\widetilde{\mathcal Q}\subset {\mathcal Q}_+$ is called ``dilationally invariant"
iff
\begin{eqnarray}\nonumber
 (Q_1,\ldots, Q_L)\in \widetilde{\mathcal Q} & \Leftrightarrow &
 (\Lambda_1 \, Q_1,\ldots,\Lambda_L\, Q_L)\in \widetilde{\mathcal Q}\quad
 \mbox{for all } \left(Q_1, \ldots,Q_L\right)\in \widetilde{\mathcal Q}\\
 \label{defdilinv}
 &&
 \mbox{ and }
 {\boldsymbol\lambda}\in {\sf D} \mbox{, where the } \Lambda_\ell \mbox{ are given by (\ref{actionQ}).}
\end{eqnarray}

Let $\widetilde{\mathcal Q}\subset {\mathcal Q}_+$ be dilationally invariant. Then a physical law of the form
$f(Q_1,\ldots, Q_L)=0$ where $f:\widetilde{\mathcal Q}\to{\mathbb R}$ will be called
{\it dilationally invariant in the domain} $\widetilde{\mathcal Q}$
iff $\{(Q_1,\ldots, Q_L)\in \widetilde{\mathcal Q}\left|f(Q_1,\ldots, Q_L)=0\right. \}$
is dilationally invariant. In most cases we will have $\widetilde{\mathcal Q}= {\mathcal Q}_+$
and the addition ``in the domain $\widetilde{\mathcal Q}$ " can be omitted.

For our proof of the $\Pi$-theorem which is inspired by \citet{C51}
it will be convenient to view  ${\mathcal Q}_+$ and  ${\mathcal R}_+$
as real, affine spaces.  A similar approach for scalar variables was pursued by \citet{D24}.
Recall that an affine space ${\mathcal A}$ is a set on which the additive group
of a vector space ${\mathcal V}$ operates transitively and freely
(for the concept of an affine space see also Section \ref{sec:NGPP}).
It follows that for any two points $A,B\in {\mathcal A}$ there is a unique vector ${\mathbf v}\in{\mathcal V}$
that maps $A$ onto $B$, and which is written as ${\mathbf v}=\stackrel{\xrightarrow{\hspace*{5mm}}}{A,B}$.
The dimension of the affine space ${\mathcal A}$ is defined as the dimension of ${\mathcal V}$.
A map $\varphi:{\mathcal A}\to{\mathcal B}$ between two affine spaces
with underlying linear space ${\mathcal V}$ and ${\mathcal W}$, resp., is called an ``affine map"
iff two conditions are satisfied. First, $\varphi$ maps parallelograms onto parallelograms, that is,
${\mathbf v}=\stackrel{\xrightarrow{\hspace*{5mm}}}{A,B}=\stackrel{\xrightarrow{\hspace*{5mm}}}{C,D}$
always implies
${\mathbf w}=\stackrel{\xrightarrow{\hspace*{1.5cm}}}{\varphi(A),\varphi(B)}=\stackrel{\xrightarrow{\hspace*{1.5cm}} }{\varphi(C),\varphi(D)}$.
This ensures that $\varphi'({\mathbf v}):={\mathbf w}$ is well-defined. The second condition
is that $\varphi':{\mathcal V}\to{\mathcal W}$ will be a linear map
(additivity of $\varphi'$ already follows from the first condition).

If ${\mathcal A}$ is an affine space with underlying vector space ${\mathcal V}$ then
each linear subspace ${\mathcal U}\subset {\mathcal V}$ gives rise to an equivalence relation
$\sim$ on ${\mathcal A}$ by means of
$A \sim B \Leftrightarrow \stackrel{\xrightarrow{\hspace*{5mm}}}{A,B}\in {\mathcal U}$.
It can be shown that the set of equivalence classes ${\mathcal A}/_\sim$ is again an affine
space with underlying vector space ${\mathcal V}/{\mathcal U}$ and that the natural
projection $\pi:{\mathcal A}\to {\mathcal A}/_\sim$ is an affine map. Moreover, every
affine map $\psi:{\mathcal A}\to{\mathcal B}$ that is constant on the equivalence classes
can be factored as $\psi=\chi \circ \pi$, where $\chi:{\mathcal A}/_\sim\to{\mathcal B}$
is a uniquely determined affine map.

We will define the action of the real vector space ${\mathbb R}^L$ on  ${\mathcal Q}_+$
by means of $\left(Q_1,\ldots,Q_L\right) \mapsto \left(e^{x_1}\,Q_1,\ldots,e^{x_L}\,Q_L\right)$
for all ${\mathbf x}=(x_1,\ldots,x_L)\in {\mathbb R}^L$ and skip the straightforward proof that this action defines
an affine structure on ${\mathcal Q}_+$. The analogous definitions apply to ${\mathcal R}_+$.
Then it can be shown that the definition (\ref{Qell}) of the derived quantities $Q_\ell$
gives rise to an affine map $\varphi: {\mathcal R}_+ \to {\mathcal Q}_+$.

We will sketch the proof of the latter.
Consider a parallelogram ${\mathbf x}=\stackrel{\xrightarrow{\hspace*{6mm}}}{R,R'}=\stackrel{\xrightarrow{\hspace*{8mm}}}{R'',R'''}$
in the affine space ${\mathcal R}_+$. This means that $R'_i=e^{x_i} R_i$ and $R'''_i=e^{x_i} R''_i$ for $i=1,\ldots,N$.
Let $Q=\varphi(R),\, Q'=\varphi(R'),\,Q''=\varphi(R''),\,Q'''=\varphi(R''')$
and ${\mathbf y}=\stackrel{\xrightarrow{\hspace*{6mm}}}{Q,Q'}$ .
From
\begin{equation}\label{QyQp}
 y_\ell =\log \Lambda_\ell \stackrel{(\ref{actionQ})}{=}\log\,\prod_{i=1}^N \lambda_i^{A_{i\ell}}
 =\sum_i A_{i\ell}\log \lambda_i
 =\sum_i A_{i\ell}x_i
 =\sum_i A^\top_{\ell i} x_i
\end{equation}
it follows that $\varphi'=A^\top$
if $\varphi'$ is well-defined. The latter follows from
\begin{equation}
\label{Qsss}
  Q'''_\ell = \prod_i \left( R'''_i\right)^{A_{i\ell}}=
  \prod_i\left( e^{x_i}\,R''_i\right)^{A_{i\ell}}=
  e^{\sum_i x_i A_{i\ell}}\,\prod_i\left(R''_i\right)^{A_{i\ell}}
  =e^{y_\ell} \,Q''_\ell
  \;,
\end{equation}
and hence ${\mathbf y}=\stackrel{\xrightarrow{\hspace*{8mm}}}{Q'',Q'''}$.
 \hfill$\Box$\\

\begin{figure}[ht!]
\centering
\includegraphics*[clip,width=0.7\columnwidth]{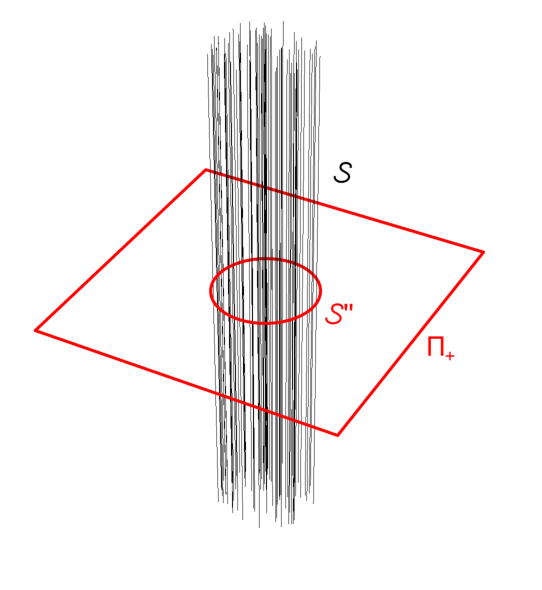}
\caption{Schematic sketch of the affine geometry of the $\Pi$ theorem.
The three-dimensional affine space shown in the figure represents ${\mathcal Q}_+$.
A dilationally invariant physical law is represented by a subset ${\mathcal S}\subset {\mathcal Q}_+$
which is invariant under translations from a certain subspace,
here represented as translations in $3$-direction.
This subspace generates an equivalence relation $\sim$ on ${\mathcal Q}_+$.
The lines parallel to the $3$-direction in the figure represent certain $\sim$ equivalence classes
contained in ${\mathcal S}$, which are generally $R$-dimensional planes in ${\mathcal Q}_+$.
Alternatively, the law can also be represented by a subset ${\mathcal S}''$ of the $K$-dimensional affine space
space $\Pi_+$, which is isomorphic to the quotient space ${\mathcal Q}_+/_\sim$.
}
\label{PIG}
\end{figure}

For the following we use the definition
\begin{equation}\label{defPiplus}
  \Pi_+= \prod_{k=1}^K {\mathbb R}_{>0}
  \;.
\end{equation}
Then the $\Pi$-theorem can be formulated as follows:\\

\begin{theorem}\label{TPi}
With the preceding notation assume that a physical law of the form
$f(Q_1,\ldots, Q_L)=0$ is dilationally invariant in the domain $\widetilde{\mathcal Q}\subset {\mathcal Q}_+$.
Then there exist $K$
dimensionless quantities with values of the form
\begin{equation}\label{Pigroup}
  \Pi_k =\prod_{\ell=1}^{L}Q_\ell^{ p_{\ell}^{(k)}},\quad \mbox{with } p_{\ell}^{(k)}\in{\mathbb Z} \mbox{ for } k=1,\ldots,K
  \;,
\end{equation}
and a function $F:\Pi_+\to {\mathbb R}$ such that
\begin{eqnarray}\nonumber
  && f(Q_1,\ldots, Q_L)=0  \,\Leftrightarrow\,  F(\Pi_1,\ldots,\Pi_K)=0     \\
   \label{PiT}
   && \mbox{for all }
 (Q_1,\ldots, Q_L)\in \widetilde{\mathcal Q} \mbox{ and }
  (\Pi_1,\ldots,\Pi_K)\in\Pi_+
  \;.
\end{eqnarray}

\end{theorem}
The proof, the details of which will be given in Appendix \ref{sec:Pi},
is based on the translation of concepts from dimensional analysis into the
language of affine geometry. A dilationally invariant law can be understood
as a collection of parallel $R$-dimensional planes and hence as a collection of
points $P$ in the corresponding $K$-dimensional affine quotient space
${\mathcal Q}_+/_\sim$, see Figure \ref{PIG}.
Each such point $P$ represents a $\Pi$-group of values of dimensionless
quantities $(\Pi_1,\ldots,\Pi_K)$ occurring in the $\Pi$-theorem.

As a simple example, let us consider the plane mathematical pendulum.
We assume that its motion is independent of its mass and we can therefore
restrict ourselves to the fundamental quantities ``length" with
the range ${\mathcal R}_1={\mathcal L}$ and ``time" with the range ${\mathcal R}_2={\mathcal T}$.
The physical parameters that belong to this system are the length
$Q_1=\ell$ of the pendulum with range ${\mathcal Q}_1={\mathcal R}_1$,
the gravitational acceleration $Q_2=g$ with range
${\mathcal Q}_2={\mathcal R}_1\otimes {\mathcal R}_2^{-2}$ and the angular frequency
$Q_3=\omega$ with range ${\mathcal Q}_3={\mathcal R}_2^{-1}$.
Thus the matrix $A$ encoding the definition of the three derived quantities
has the form
\begin{equation}\label{defA}
 A=\left(
 \begin{array}{ccc}
   1 & 1 & 0 \\
   0 & -2 & -1
 \end{array}
 \right)
 \;.
\end{equation}
Its kernel is one-dimensional and spanned by the vector $(1,-1,2)\in {\mathbb Z}^3$.
This means that the $\Pi$-group of dimensionless quantities consists of a single element
of the form
\begin{equation}\label{PigroupEx}
\Pi_1= Q_1^{1}\,Q_2^{-1}\,Q_3^{2}= \frac{\ell}{g}\omega^2
\;.
\end{equation}
The law therefore has the form $0=F(\Pi_1)=F(\frac{\ell}{g}\omega^2)$ which is satisfied by the equation
\begin{equation}\label{omegasqrt}
 \omega =\sqrt{\frac{g}{\ell}}    \quad \Leftrightarrow\quad  F\left(\frac{\ell}{g}\omega^2\right)= \frac{\ell}{g}\omega^2-1=0
\end{equation}
for the angular frequency of small oscillations of the pendulum about its rest position.

One may ask why only the form of the approximation law for small oscillations
is obtained by the $\Pi$-theorem and not the form of the
correct law for oscillations with finite angular amplitudes $\theta_0$.
The answer is that you cannot expect miracles.
If $\theta_0$ is another important parameter for the problem, you have to tell the $\Pi$-theorem.
Its answer would then be that the new $\Pi$-group consists of two elements,
the already known $\Pi_1$ and the second dimensionless quantity $\Pi_2=\theta_0$. Hence the
new form of the law will be $F(\frac{\ell}{g}\omega^2, \theta_0)=0$.
As one can easily see, the explicit law for the angular frequency of oscillations with finite amplitude
\begin{equation}\label{omegaexact}
  \omega =\sqrt{\frac{g}{\ell}}\,\frac{\pi/2}{K(\sin \frac{\theta_0}{2})}
  \;,
\end{equation}
where $K$ is the complete elliptic integral of the first kind, see \citep[\S 19.2(ii)]{N23},
can be reduced to this form.

\subsection{$3$-dimensional vector quantities}\label{sec:VQ}

For the second example of a physical theory involving dimensional quantities,
which will be presented in Section \ref{sec:NGPP},
we need an extension of the previous considerations to $3$-dimensional vector quantities.
There are plenty of these vector quantities in physics,
just to mention velocity, acceleration, force, angular momentum, electric and magnetic fields.

We will define the notion of a $3$-dimensional real Euclidean vector space $V$
over the range of (fundamental or derived) scalar quantities ${\mathcal V}={\mathcal R}^{({\mathbf n})}$, the latter equipped
with a positive range ${\mathcal V}_{>0}\subset{\mathcal V}$.
To this end we adopt the usual definition of a $3$-dimensional real vector space $V$.
Further we assume a map
\begin{equation}\label{defscalar}
\langle\,,\, \rangle : V\times V \rightarrow {\mathcal V}\otimes {\mathcal V}
\;,
\end{equation}
and postulate axioms analogously to those of a Euclidean scalar product.
This means, that $\langle\,,\, \rangle$ is required to be bilinear, symmetric and
to satisfy $\langle {\mathbf v},{\mathbf v}\rangle\in {\mathcal V}_{>0}\otimes {\mathcal V}_{>0}$ for all
${\mathbf v}\in V,\; {\mathbf v}\neq {\mathbf 0}$
and that $\langle {\mathbf v},{\mathbf v}\rangle=0$ implies $ {\mathbf v}={\mathbf 0}$.

Let ${\mathcal U}$ be the range of another (fundamental or derived) scalar quantity and  ${\mathcal U}_{>0}$ its
positive range. Then it is possible to multiply the vectors ${\mathbf v}\in V$ with
this quantity and thus to obtain another  $3$-dimensional real Euclidean vector space $W$
over the scalar range ${\mathcal W}={\mathcal V}\,{\mathcal U}$.
To show this we will define $W$ as the $3$-dimensional vector spaces
$W=V\otimes {\mathcal U}$ and the
scalar product between vectors of  $W$ as
\begin{eqnarray}\label{defscalarW1}
 \langle\,,\, \rangle_W & : & W\times W \rightarrow {\mathcal W}\otimes{\mathcal W};,\\
\label{defscalarW2}
\langle{\mathbf w}_1,{\mathbf w}_2 \rangle_W
&:=&
\langle{\mathbf v}_1, {\mathbf v}_2 \rangle_V\;U_1\,U_2
\;,
\end{eqnarray}
where ${\mathbf w}_i={\mathbf v}_i\otimes U_i$ for $i=1,2$. The r.~h.~s.~of (\ref{defscalarW2})
has to be understood as the product of a quantity with range ${\mathcal V}^2$ and
another quantity with range ${\mathcal U}^2$ resulting in a quantity
with range ${\mathcal W}^2$.
The norm $\|{\mathbf v}\|$ of a vector ${\mathbf v}\neq {\mathbf 0}$ is defined as
\begin{equation}\label{defnorm}
 \|{\mathbf v}\|:=\sqrt{\langle {\mathbf v}\left|\right.{\mathbf v}\rangle  }
 \;,
\end{equation}
where the square root of a scalar quantity has been defined in (\ref{defsqr}) and yields a value in ${\mathcal V}_{>0}$
if ${\mathbf v}\neq {\mathbf 0}$. For  ${\mathbf v}={\mathbf 0}$ one defines $\|{\mathbf v}\|:=0\in {\mathcal V}$.

These definitions are almost identical to those in \citep[Sections 5 and 6]{D24},
except for the (insignificant) error in the definition of the range of the scalar product
in \citep[Eq.~(20)]{D24}, which can easily be corrected,
for example by using the methods outlined in Appendix \ref{sec:ACP}.

As an application of the preceding definitions we consider the usual
decomposition of a non-zero vector ${\mathbf v}\in V$ in the form
\begin{equation}\label{dirmag}
 {\mathbf v}=\underbrace{\frac{\mathbf v}{\|{\mathbf v}\|}}_{\mbox{direction}}\;\cdot\; \underbrace{\|{\mathbf v}\|}_{\mbox{magnitude}}
 \;.
\end{equation}
We can interpret the direction $\frac{\mathbf v}{\|{\mathbf v}\|}$ as a vector quantity
with values in ${\mathcal V}^{-1}\, V$ and the magnitude as a scalar quantity with values in ${\mathcal V}_{>0}$.
The vector space ${\mathcal V}^{-1}\, V$  is the dimensionless version of $V$ and its scalar product
according to the definition (\ref{defscalarW2}) is real-valued and Euclidean in the proper sense.

Consider an orthogonal basis $\left( {\mathbf e}_1,{\mathbf e}_2,{\mathbf e}_3\right)$ in $V$
such that all basis vectors have the same length $\ell\in {\mathcal V}_{>0}$.
Then each vector ${\mathbf v}\in V$ has the unique representation
${\mathbf v}=\sum_{i=1}^{3} v_i\,{\mathbf e}_i$ with real numbers $v_i,\,i=1,2,3$.
From this we can derive the following representation
\begin{equation}\label{repv}
  {\mathbf v}=\sum_{i=1}^{3} \underbrace{v_i\,\ell}_{\in {\mathcal V}}\quad
  \underbrace{\frac{{\mathbf e}_i}{\ell}}_{\in {\mathcal V}^{-1}\, V}
  \;,
\end{equation}
which uses the corresponding orthonormal basis in ${\mathcal V}^{-1}\, V$.
This is consistent with the usage in physics to represent
a $3$-vector with physical dimension as a triple of values of a scalar quantity.

\section{Species of structure}\label{sec:SS}
\subsection{General account}\label{sec:SSGA}

We will only recall the essential ingredients of ``species of structure"
using informal set theory. For more detailed accounts
see  \citep[section 7]{L90},  \citep[section 1.3]{BMS87}, \citet{S82}
and \citep{B04}.
A species of structure $\Sigma$ is given by the quintuple $\Sigma=\langle A,B,\tau,\sigma,\alpha\rangle$.
Here $A$ and $B$ represent finite collections of {\em base sets}, where
the base sets of $A=\langle A_1,\ldots,A_a\rangle$ are called ``auxiliary base sets"
and those of $B=\langle B_1,\ldots,B_b\rangle$ ``principal base sets".
For the difference between these two notions see below.

From base sets one can construct new sets using the Cartesian product
and the power set (set of all subsets of a set).
By recursively applying these constructions
one obtains so-called ``echelon sets" $\tau(A,B)$,
which are formed according to the ``echelon construction scheme" $\tau$.
The ``structural element" $\sigma$ is an element of such an echelon set, i.~e.,
$\sigma\in \tau(A,B)$.
Without loss of generality we may assume that $\sigma$ has $S$ components, and thus of the
form $\sigma=(\sigma_1,\ldots, \sigma_S)\in \tau_1(A,B)\times\ldots \times \tau_S(A,B)$.

We will present physical examples below, but for the moment only mention
two examples: First, a topological space $B_1$ (no auxiliary base sets) which
can be characterized by the set of open subsets
$\sigma \in\tau(B_1)={\mathcal P}\left({\mathcal P}\left(B_1\right) \right)$.
As a second example, an ordered set $B_1$ has the structural element $<$
that is a relation on $B_1$, i.~e.,
$<\;\in \tau(B_1)= {\mathcal P (B_1\times B_1)}$.

The final component of $\Sigma$ is the {\em axiom} $\alpha=\alpha(A,B,\sigma)$
which in physical applications includes the laws of the theory.
There is some postulate for $\alpha(A,B,\sigma)$ called ``transportability".
It essentially says that $\alpha(A,B,\sigma)$ characterizes the
mathematical objects $B$ and $\sigma$ only up to isomorphisms.
More precisely, we consider an $a+b$-tuple of bijections
\begin{equation}\label{aplusbtuple}
 f=\langle {\rm id},\ldots,{\rm id}, f_1,\ldots,f_b\rangle,
 \quad\mbox{where}\; f_i:A_i\rightarrow A_i'\;\mbox{for all}\;i=1,\ldots,b
 \;,
\end{equation}
and define $\sigma^f:= \tau(f)(\sigma)$, where $\tau(f)$ denotes the canonical extension of $f$ to the
echelon set $\tau(A,B)$ and ${\rm id}$ stands for the identity mappings ${\rm id}:A_i\rightarrow A_i,\;i=1,\ldots,a$.
The transportability postulate then says that $\alpha(A,B,\sigma)\Leftrightarrow \alpha(A,B',\sigma^f)$
for all $f$ and $B'$ with the above properties.
This is the only place in the definition of a species of structure where a distinction
between auxiliary base sets and principal base sets is made. In physical applications
the auxiliary base sets are often fixed sets like ${\mathbb R}$ or ${\mathbb C}$ together with their structural components
and axioms. But in  \citep{B04} also the case of a species of structure
for a vector space $V$ over a field $F$ is mentioned, where $V$ is a principal base set
and $F$ is an auxiliary one. This has the consequence that automorphisms of $V$ leave
the field $F$ fixed which makes sense. For the physical interpretation of the transportability postulate
see also  \citet{S91}, \citet{S94}.

For the definition of an automorphism of $\Sigma$ we consider again an $a+b$-tuple of bijections
\begin{equation}\label{aplusbtuple}
 f=\langle {\rm id},\ldots,{\rm id}, f_1,\ldots,f_b\rangle,
 \quad\mbox{where}\; f_i:A_i\rightarrow A_i\;\mbox{for all}\;i=1,\ldots,b
 \;,
\end{equation}
and say that $f$ is an {\em automorphism} of $(A,B,\sigma)$ iff $\sigma^f=\sigma$,
in symbols: $f\in \mbox{Aut}(A,B,\sigma)$.
The set of automorphisms of fixed $(A,B,\sigma)$ forms a group. It may vary with $(A,B,\sigma)$.
Only if every pair $(A,B_i,\sigma_i), i=1,2,$ satisfying $\alpha$ is isomorphic
(in this case the species of structure is called {\it categorical}) then the
automorphism group of $\Sigma$ is unique (up to isomorphism).

As an example, we consider the species of structure $\Sigma O$,
which belongs to the theory $TO$ described in section \ref{sec:Ohm}.
There are certain ``stylistic differences"' between the various structuralist schools
in the use of species of structure $\Sigma$ in the reconstruction of physical theories.
In the Sneed school, roughly speaking, individual experiments are conceptualized as models of
$\Sigma$ and relationships between experiments are introduced as ``constraints".
We tend to follow the Ludwig school here, which also allows sets of experiments as base sets
of $\Sigma$ such that ``constraints"  can be captured as part of the structural element of $\Sigma$ or its axiom.

The only auxiliary base set of $\Sigma O$ will be $A_1={\mathbb R}$.
There is a corresponding component $\sigma_{\mathbb R}$ of the structural element of $\Sigma O$
and a corresponding part $\alpha_{\mathbb R}$ of its axiom, which we will assume to be well known.
The structural element will be of the form $\sigma = \left( \sigma_{\mathbb R},\sigma_U,\sigma_I,\sigma_B\right)$
and the axiom of $\Sigma O$ is written as $\alpha=\alpha_{\mathbb R}\wedge \alpha_U\wedge\alpha_I\wedge\alpha_B$
to be explained in what follows.
The principal base sets will be $B=\langle {\mathcal U},{\mathcal I},{\sf R}\rangle$.
Recall that ${\mathcal U}$ and ${\mathcal I}$ are one-dimensional, real, ordered vector spaces.
This requires some structural components $\sigma_U\in \tau_U(A,B)$ and $\sigma_I\in \tau_I(A,B)$
and corresponding axioms $\alpha_U$ and  $\alpha_I$, which we will not give in details, but only
mention that
\begin{equation}\label{strucU}
 \sigma_U=\left({\mathcal U}_{>0}, +, \cdot\right)\in\tau_U(A,B):=
  {\mathcal P} {\mathcal U}\times{\mathcal P} \left( {\mathcal U}\times {\mathcal U}\times{\mathcal U}\right)
  \times
  {\mathcal P} \left( {\mathbb R}\times {\mathcal U}\times{\mathcal U}\right)
  \;.
\end{equation}
The central structural component will be
\begin{equation}\label{gamma}
 \sigma_B:= \gamma \in \tau_B(B):={\mathcal P}\left({\sf R}\times {\mathcal U}\times {\mathcal I}\right)
  \;,
\end{equation}
corresponding to the axiom $\alpha_B$  which is given by the Axioms $TO$ \ref{Axiom1} and $TO$ \ref{Axiom2}.

This concludes the description of $\Sigma O$,
in which many details have been omitted, but which the reader can still easily reconstruct.
Noteworthy, the automorphism group of
$\left({\mathbb R},{\sf R},{\mathcal U},{\mathcal I};\sigma\right)$ contains
the two-dimensional group of positive dilations $(\lambda,\mu)$ of ${\mathcal U}$ and ${\mathcal I}$.
This subgroup operates on resistors $\rho\in{\sf R}$ by means of
a bijection $f_1: {\sf R}\to {\sf R}$ satisfying
${\sf m}(f_1(\rho))=\frac{\lambda}{\mu}{\sf m}(\rho)$,
where ${\sf m}:{\sf R}\to {\mathcal R}_{>0}$ was the map that assigns to each resistor
$\rho$ its resistance ${\sf m}(\rho)$, see Section \ref{sec:Ohm}.
Automorphism groups can be interpreted physically in a passive or active sense,
as explained in more detail in section \ref{sec:AP}.
In the passive interpretation, the positive dilations $(\lambda,\mu)$
can be understood as a change in the physical units for voltage and current.
In the active interpretation the positive dilation $(\lambda,\mu)$
map a possible experiment $(\rho,U,I)$ onto another one with outcome
$\left(f_1(\rho),\lambda\,U,\mu\,I\right)$ and hence
represents some kind of similarity transformation of Ohm's law.

\subsection{Species of structure and $\Pi$-theorem}\label{sec:SSPT}

Returning to the general case, we will examine how the $\Pi$-theorem
fits into the framework of species of structures.
In particular, we will take a look at the connection between the dilation invariance of physical laws and
automorphisms of a modified species of the structure $\Sigma_0$.

In general, the physical law dealt with in the $\Pi$-theorem will not be the
complete axiom $\alpha(A,B,\sigma)$, but only a logical consequence of $\alpha$,
i.~e.~a theorem in the theory of $\Sigma$, which will be denoted by $\alpha_0(A,B,\sigma_0)$.
Here, $\sigma_0$ is another structural term with the typification $\sigma_0\in {\mathcal P}\tau_0(A,B)$
where $\tau_0$ is an echelon construction scheme.
Formally, this becomes a modified species of structure $\Sigma_0$ with the same base sets
$A,B$ as $\Sigma$ if it can be ensured that $\alpha_0(A,B,\sigma_0)$ again will be transportable.

Moreover, the ranges of the fundamental quantities ${\mathcal R}_i,\,i=1,\ldots, N,$
are assumed to be among the principal base sets $B_j,\,j=1,\ldots,b,$ of $\Sigma$.
As in subsection \ref{sec:BPT} we consider the derived quantities with ranges
${\mathcal Q}_\ell=\bigotimes_{i=1}^N  {\mathcal R}_i^{A_{i\ell}},\,\ell=1,\ldots,L$.
Further, some maps $q_\ell: \tau_0(A,B) \to {\mathcal Q}_\ell,\; \ell=1,\ldots,L,$
are defined such that $\alpha_0(A,B,\sigma_0)$ will be equivalent to
\begin{equation}\label{alpha0}
\sigma_0=\{{\mathbf y}\in \tau_0(A,B)\left|F\left(q_1({\mathbf y}),\ldots,  q_L({\mathbf y})\right)=0\right.\}
 \;,
\end{equation}
with some function $F:{\mathcal Q}_+\to {\mathbb R}$.
(we avoid the letter ``$f$" which already denotes the automorphism).
Let
\begin{equation}\label{deftildeQ}
 \widetilde{\mathcal Q}:= \{\left( q_1({\mathbf y}),\ldots,q_L({\mathbf y})\right)\left| {\mathbf y}\in  \tau_0(A,B)\right. \}
 \;.
\end{equation}

It is a standard assumption that all terms defined in the theory of $\Sigma$ are
``intrinsic", i.~e., transform covariantly under isomorphisms,
see \citet{B04} or \citet{L90} for more details.
For the special case of an automorphism $f$ it follows that an intrinsic term
will be invariant under $f$.
Hence we will also assume that the maps $q_\ell,\,\ell=1,\ldots,L$ have this property.

In particular, we consider automorphisms $f=(\mbox{id },\ldots,\mbox{id}, f_1,\ldots, f_b)$
of $(A,B,\sigma_0)$ that
act as positive dilations $R_i \mapsto \lambda_i\,R_i$
on the ranges of fundamental quantities ${\mathcal R}_i,\;i=1,\ldots,N,$
and in a suitable way onto other principal base sets.
Then the assumption that the maps $q_\ell,\,\ell=1,\ldots,L$ will be invariant under $f$
can be written as
\begin{equation}\label{qintrinsic}
q_\ell( {\mathbf y}^f)= \prod_{i=1}^{N}\lambda_i^{A_{i\ell}}\, q_\ell( {\mathbf y})
\stackrel{(\ref{actionQ})}{=}\Lambda_\ell\,  q_\ell( {\mathbf y})
\quad \mbox{for all }\ell=1,\ldots,L \mbox{ and } {\mathbf y}\in \tau_0(A,B)
\;.
\end{equation}
It follows that $\widetilde{\mathcal Q}$ defined in (\ref{deftildeQ}) will be dilationally invariant.

With these preparations we can formulate the following
\begin{prop}\label{P1}
  Assume that for all ${\boldsymbol\lambda}\in{\sf D}$ there exists an automorphism
  $f=(\mbox{id },\ldots,\mbox{id}, f_1,\ldots, f_b)$ of $(A,B,\sigma_0)$
 that acts in the form of positive dilations $R_i \mapsto \lambda_i\,R_i$
on the ranges of fundamental quantities ${\mathcal R}_i,\;i=1,\ldots,N$.
Then the physical law given by (\ref{alpha0}) will be dilationally invariant in the domain $\widetilde{\mathcal Q}$.
\end{prop}
{\bf Proof}: Let $g=f^{-1}$ and consider the following identities:
\begin{eqnarray}
\label{proofPa}
\sigma_0&\stackrel{(\ref{alpha0})}{=}& \{{\mathbf y}\in \tau_0(A,B)\left|F\left(q_1({\mathbf y}),\ldots,  q_L({\mathbf y})\right)=0\right.\}\\
 \label{proofPb}
  &=&\sigma_0^g = \{{\mathbf y}^g\in \tau_0(A,B)\left|F\left(q_1({\mathbf y}),\ldots,  q_L({\mathbf y})\right)=0\right.\} \\
  \label{proofPc}
  &=& \{{\mathbf y}\in \tau_0(A,B)\left|F\left(q_1({\mathbf y}^f),\ldots,  q_L({\mathbf y}^f)\right)=0\right.\}\\
  \label{proofPd}
   &\stackrel{(\ref{qintrinsic})}{=}& \{{\mathbf y}\in \tau_0(A,B)
   \left|F\left(\Lambda_1\,q_1({\mathbf y}),\ldots, \Lambda_L\, q_L({\mathbf y})\right)=0\right.\}
   \;,
\end{eqnarray}
where (\ref{proofPb}) holds since with $f$ also $g=f^{-1}$ is an automorphism of ${(A,B,\sigma_0)}$.
From this we conclude
\begin{equation}\label{proofPe}
 F\left(q_1({\mathbf y}),\ldots, q_L({\mathbf y})\right)=0 \;\Leftrightarrow\;
 F\left(\Lambda_1\,q_1({\mathbf y}),\ldots, \Lambda_L\, q_L({\mathbf y})\right)=0
\end{equation}
for all ${\mathbf y}\in \tau_0(A,B)$ and ${\boldsymbol\lambda}\in{\sf D}$.
Hence the physical law given by (\ref{alpha0}) will be dilationally invariant in the domain $\widetilde{\mathcal Q}$.
\hfill$\Box$\\

It seems appropriate to explain this proposition for the example of Ohm's theory $TO$.
Here we choose $\sigma_0=\sigma_B=\gamma$ according to (\ref{Gamma}) and Axiom \ref{Axiom2}
such that $\tau_0(A,B)={\sf R}\times{\mathcal U}\times {\mathcal I}$.
Further we consider the automorphism group of $\left({\mathbb R},{\mathcal U},{\mathcal I},{\sf R};\sigma\right)$
described at the end of subsection \ref{sec:SSGA} involving positive dilations $U\mapsto \lambda\, U$ and $I\mapsto \mu\,I$.
$\gamma \subset {\sf R}\times{\mathcal U}_{>0}\times {\mathcal I}_{>0}$ will be invariant under
these automorphisms. The physical law $U={\sf m}(\rho)\,I$ is of the above form
$\sigma_0= \{{\mathbf y}\in \tau_0(A,B)\left|F\left(q_1({\mathbf y}),\ldots,  q_L({\mathbf y})\right)=0\right.\}$
if we choose $q_1={\sf R}\times{\mathcal U}\times {\mathcal I}\to {\mathcal R},\; q_1(\rho,U,I)={\sf m}(\rho)$,
$q_2:{\sf R}\times{\mathcal U}\times {\mathcal I}\to {\mathcal U},\; q_2(\rho,U,I)=U$ and
$q_3:{\sf R}\times{\mathcal U}\times {\mathcal I}\to {\mathcal I},\; q_3(\rho,U,I)=I$.
Moreover, $F({\sf m}(\rho),U,I)$ will be chosen as
$F({\sf m}(\rho),U,I)=\frac{{\sf m}(\rho)\,I}{U}-1$ which is already in dimensionless form. Here we have
again used the identification of $\mbox{Lin}\left( {\mathcal I}\to {\mathcal U}\right)$
with the range  ${\mathcal R}={\mathcal U}\otimes{\mathcal I}^\ast$.

\section{Theory of $N$ gravitating point particles}\label{sec:NGPP}
\subsection{General account}\label{sec:NGGA}

In this section we will apply the formalism of physical quantities developed so far
to a physical theory that is still simple but no longer has the character of a toy theory,
such as the theory of Ohm's law considered in Section \ref{sec:Ohm}.
This will be the Newtonian theory $TN$ of $N$ gravitating point particles.
A similar reconstruction of $TN$ using an account of dimensional quantities
has recently published by \citet{D24}. We will point out some differences to
our approach below.
In order not to burden the discussion with topics that are not at issue here,
we do not consider Galilean spacetime for this theory,
as would actually be ``state of the art", but space and time separately.
We also assume that the masses of the $N$ point bodies are already
known from other measurement methods, whereby we do not distinguish
between gravitational and inertial masses. This assumption would be violated
in celestial mechanics, for example, in which the mass of the planets
can only be measured by applying Newton's law of gravitation and would
therefore have the status of a "theoretical quantity".
But, as said before, we want to avoid this topic here.

For $TN$ we assume three ranges of fundamental physical quantities,
${\mathcal L},\; {\mathcal T}$ and ${\mathcal M}$, for, resp., ``length", ``time", and ``mass".
The physical position space is assumed to be a $3$-dimensional, real, affine space ${\sf A}$ over
the $3$-dimensional Euclidean vector space $V$ of dimension ``length".
It is appropriate to explain these terms a little.

The affine space ${\sf A}$ over the vector space $V$ is a set of points
on which the additive group $(V,+)$ operates {\em transitively} and {\em freely}
by means of a map $\mbox{op}:V\times {\sf A}\to {\sf A}$.
We will denote the group action $\mbox{op}({\mathbf v},{\mathbf a})$ as
${\mathbf v}+{\mathbf a}$ for ${\mathbf v}\in V$
and  ${\mathbf a}\in {\sf A}$. Then ``transitive action"
means that for every ${\mathbf a},{\mathbf b}\in {\sf A}$
there exists a  ${\mathbf v}\in V$ such that ${\mathbf v}+{\mathbf a}={\mathbf b}$.
The action is ``free" iff ${\mathbf v}+{\mathbf a}={\mathbf a}$ for some ${\mathbf a}\in {\sf A}$
already implies ${\mathbf v}={\mathbf 0}$. In this case the equation
${\mathbf v}+{\mathbf a}={\mathbf b}$ uniquely determines ${\mathbf v}$
and we will write ${\mathbf v}={\mathbf b}-{\mathbf a}$. Upon fixing an ``origin"
${\mathbf 0}\in {\sf A}$, the affine space ${\sf A}$ can be identified with $V$ by means of
${\mathbf v}={\mathbf b}-{\mathbf 0}$,
but it would be unfounded to fix such an origin for the reconstruction of the theory $TN$.
If  $V$ is a topological vector space, which implies that its topology $\tau_V$ is
translationally invariant, then this topology can be transferred to ${\sf A}$
by means of the above-mentioned identification. This topology on ${\sf A}$
is independent of the choice of ${\mathbf 0}\in {\sf A}$ due to the
translational invariance of $\tau_V$.
The above applies in particular to the standard topology of finite-dimensional
vector spaces $V$, which are considered here.

We further assume that $V$ is a $3$-dimensional, real, Euclidean vector space
with dimension ${\mathcal L}$, in the sense defined in Section \ref{sec:VQ}.
Analogously, the time axis will be represented in $TN$ by a one-dimensional
affine space $\sf{T}$ over the one-dimensional Euclidean
vector space $T$ with dimension ${\mathcal T}$.
The question why not identify $T$ and ${\mathcal T}$ will be addressed below.
Another basic quantity is ${\sf P}$, the set of all particles
that could be involved in gravitational interactions.
In order to obtain concise results for the group of automorphisms it will be
convenient to identify particles of the same mass.
Concrete systems of interacting particles
always consist of a finite number of particles, some of which may have the same mass.
Hence they will be characterized by maps $P:\{1,2,\ldots,N\}\to{\sf P}$ with integers $N>1$
which are not required to be injective.

We will proceed by considering the traditional form of the equation of motion
of the gravitating particles and then discuss the re-interpretation of this
using the present formalism. The equation of motion reads
\begin{equation}\label{eom1}
  m_i\, \frac{d^2}{dt^2}{\mathbf x}_i(t)= \Gamma \sum_{j=1,j\neq i}^{N} m_i\,m_j\;
  \frac{{\mathbf x}_j(t)-{\mathbf x}_i(t)}{\left|{\mathbf x}_j(t)-{\mathbf x}_i(t) \right|^3}
  \;,
\end{equation}
for all $i=1,\ldots, N$, with the masses $m_i$ of the $N$ particles, their positions ${\mathbf x}_i(t)$
at time $t$, and the gravitational constant $\Gamma$.
Since we have not distinguished between gravitational and inertial mass we may cancel
the $m_i$ on both sides of (\ref{eom1}) and obtain
\begin{equation}\label{eom2}
 \frac{d^2}{dt^2}{\mathbf x}_i(t)= \Gamma \sum_{j=1,j\neq i}^{N} m_j\;
  \frac{{\mathbf x}_j(t)-{\mathbf x}_i(t)}{\left|{\mathbf x}_j(t)-{\mathbf x}_i(t) \right|^3}
  \;.
\end{equation}

Now we come to the reconstruction of $TN$ and re-interpretation of (\ref{eom2}).
The structural element of the species of structure $\Sigma N$ pertaining to $TN$ includes
the structural components $m$ and ${\mathbf X}$. Here $m$ is the map
\begin{equation}\label{mapm}
 m: {\sf P}\rightarrow {\mathcal M}_{>0}
 \;,
\end{equation}
which will be assumed to be bijective according to the above-mentioned identification of particles with the same mass.
The mass values $m_i$ occurring in (\ref{eom2}) have to be read\\ as $m_i = (m\circ P)(i),\, i=1,\ldots,N$.

Further, ${\mathbf X}$ will be a set of pairs $(P,{\mathbf x})$ where
$P:\{ 1,2,\ldots,N\}\to {\sf P}$ as explained above
and ${\mathbf x}$ is a map of the form
\begin{equation}\label{mapm}
{\mathbf x}: \{1,2,\ldots, N\} \times (a,b)\rightarrow A
 \;,
\end{equation}
which defines the position ${\mathbf x}_i(t)\in A$ of the $i$-th particle at time $t\in (a,b)$.
The finite set $\{1,2,\ldots, N\}$ can be understood as a subset of the auxiliary base set $A_1={\mathbb R}$.
Moreover, $(a,b)\subset  \sf{T}$ is an open interval of the one-dimensional affine space $\sf{T}$, where
the special cases $a=-\infty$ or $b=\infty$ are also possible. This formulation seems to depend on a chosen orientation of
$\sf{T}$ but could be understood independently of any orientation.
Intuitively, ${\mathbf X}$ will represent the set of solutions of the equations of motion of the form (\ref{eom2}).
A single solution $(P,{\mathbf x})\in {\mathbf X}$ can be thought of a single gravitational experiment
performed on the system of particles $P$.
The restriction to  $t\in (a,b)$ is necessary since two particles may collide at finite times such that the
r.~h.~s.~of (\ref{eom2}) is no longer defined.
The advantage of the restriction $t\in(a,b)$ is a mathematically correct form of the
central axiom of TN, even if this may seem somewhat pedantic.
If we assume that at some point in time $t$ all $N$ point particles have different positions,
then this also applies to an open interval $(a_0,b_0)$ including $t$
since the solutions of  (\ref{eom2}) depend continuously on $t$.
Either the end point $b_0$ of this interval can be extended indefinitely,
or there is a limit value $b$ at which at least two particles collide.
The same applies to $a_0$. In this way, we obtain a maximal interval $(a,b)$
where the equation of motion is well-defined, see also Axiom $TN$ \ref{AxTN} below.
The typification of ${\mathbf X}$ according to the given explanations will be denoted by
${\mathbf X}\in \sigma_X= {\mathcal P}\left(\sigma_x \right)$.

We now turn to the interpretation of the l.~h.~s.~of (\ref{eom2})
and consider the first derivative $\frac{d}{dt}{\mathbf x}_i(t)$.
It turns out that the definition of $\frac{d}{dt}{\mathbf x}_i(t)$
depends on the choice of the orientation of $T$. $T$ is a one-dimensional
vector space with two different orientations (``arrows of time") represented by subsets of
positive elements $T_{>0}^{(i)},\; i=1,2$, but there is no physical reason to
single out one of these orientations in the theory $TN$. In contrast,
${\mathcal T}$ was assumed to be ordered, since, after all, the unit of time should be positive.
This difference between $T$ and ${\mathcal T}$ is an argument
for distinguishing between these two vector spaces in the reconstruction of $TN$.
Only then can the question of whether $TN$ is invariant under time reflections be investigated at all.

Let $t\in{\sf{T}}$ be fixed and consider some variable $\epsilon\in  T_{>0}^{(i)},\; i=1,2$.
Since $T$ operates on ${\sf{T}}$, we have $\epsilon+t\in {\sf{T}}$ and
${\mathbf x}_i(\epsilon+t)-{\mathbf x}_i(t)\in V$. This vector
has to divided by the quantity
\begin{equation}\label{abseps}
  \left|\epsilon\right| := \sqrt{\langle  \epsilon,\epsilon\rangle}
\end{equation}
of dimension ${\mathcal T}$ defined by (\ref{defsqrt}).
The product
$\left({\mathbf x}_i(\epsilon+t)-{\mathbf x}_i(t)\right) \frac{1}{ \left|\epsilon\right|}$
will be understood in the sense of subsection \ref{sec:VQ} as the product of a $3$-dimensional
vector quantity of dimension ${\mathcal L}$ with the derived scalar quantity of dimension ${\mathcal T}^{-1}={\mathcal T}^\ast$.
Hence it is a $3$-dimensional vector quantity of dimension  ${\mathcal L}\otimes {\mathcal T}^\ast$,
called ``velocity". In the limit $\epsilon\to 0$ we obtain the vector quantity
denoted by $\frac{d}{dt}{\mathbf x}_i(t)$. Of course, it has to be postulated in the axioms of $TN$
that this limit exists for all $t\in (a,b)$ and $i=1,\ldots,N$, in other words, that
${\mathbf x}_i(t)$ is differentiable w.~r.~t.~$t$.
As we have noted, the definition of $\frac{d}{dt}{\mathbf x}_i(t)$ depends on the
choice of the orientation of $T$. The change from one orientation to the other
produces a minus sign in $\frac{d}{dt}{\mathbf x}_i(t)$,
in accordance with the textbook wisdom of how velocity transforms under time reflection.

Analogously, we can treat the second derivative and obtain
$\frac{d^2}{dt^2}{\mathbf x}_i(t)$ as a $3$-dimensional vector quantity of dimension
${\mathcal L}\otimes {\mathcal T}^\ast\otimes {\mathcal T}^\ast={\mathcal L}\,{\mathcal T}^{-2}$,
called ``acceleration".
Changing the orientation of $T$ results in a further minus sign
for the derivative of the velocity, and thus the acceleration is independent of the orientation of $T$.

Next we consider the r.~h.~s.~of (\ref{eom2}). The term ${\mathbf x}_j(t)-{\mathbf x}_i(t)$
can easily be understood as a $3$-dimensional vector quantity of dimension ${\mathcal L}$.
The term $\left|{\mathbf x}_j(t)-{\mathbf x}_i(t) \right|$ in the denominator has to be read as
\begin{equation}\label{denrhs}
  \left|{\mathbf x}_j(t)-{\mathbf x}_i(t) \right| =
  \sqrt{\langle {\mathbf x}_j(t)-{\mathbf x}_i(t) ,{\mathbf x}_j(t)-{\mathbf x}_i(t) \rangle}
  \;,
\end{equation}
in the sense of (\ref{defsqrt}), and is hence a quantity of dimension ${\mathcal L}$.
Consequently, the terms in the sum of (\ref{eom2}) are $3$-dimensional vector
quantities of dimension ${\mathcal L}$ multiplied by $m_j\,\left|{\mathbf x}_j(t)-{\mathbf x}_i(t) \right|^{-3}$,
a scalar quantity of dimension ${\mathcal M}{\mathcal L}^{-3}$.  Due to the cancelling operator ${\sf C}$
built in the definition of multiplication of quantities, the r.~h.~s.~of (\ref{eom2})
is hence the product of a  $3$-dimensional vector quantity of dimension ${\mathcal M}{\mathcal L}^{-2}$
times a quantity $\Gamma$.

It remains to formulate the main axiom of $TN$ that comprises (\ref{eom2}).
Adopting the interpretation of both sides of (\ref{eom2}) given above we may
formulate this axiom as follows:\\

\begin{axN}\label{AxTN}
 There exists a positive scalar quantity $\Gamma$ of dimension
 ${\mathcal L}^{3}{\mathcal T}^{-2}{\mathcal M}^{-1}$ such that for all
 $(P,{\mathbf x})\in{\sigma_x}$  the following holds:
 $(P,{\mathbf x})\in{\mathbf X}$ iff
 there exists an integer $N>1$ and an open interval $(a,b)\subset {\sf T}$ such that
 \begin{enumerate}
   \item $P$ is a map $P:\{1,2,\ldots,N\}\to{\sf P}$,
   \item ${\mathbf x}$ is a map ${\mathbf x}:\{1,2,\ldots,N\}\times (a,b) \to {\sf A}$,
   that is twice continuously differentiable w.~r.~t.~the second argument,
   \item the positions  ${\mathbf x}(i,t)$ and  ${\mathbf x}(j,t)$ are different for all $1\le i\neq j\le N$ and $t\in(a,b)$,
   \item the l.~h.~s.~of (\ref{eom2}) and the r.~h.~s.~of (\ref{eom2}) agree, and
   \item the time interval $(a,b)$ is maximal.
 \end{enumerate}
\end{axN}
Here we adopt the definition that $(a,b)$ is maximal iff, either $a=\infty$ or,
for finite $a$, $\lim_{t\to a} {\mathbf x}(i,t) =\lim_{t\to a} {\mathbf x}(j,t)$
for at least one pair $i,j\in P,\; i\neq j$ and analogously for $b$.

The gravitational constant $\Gamma$ has been treated as a ``theoretical quantity",
it is not part of the structural term of $T\,N$ but occurs in the axiom $TN$ \ref{AxTN}
of the theory as a variable bounded by an existential quantifier.
Theoretical quantities can only be measured by applying the theory in which they occur,
This is a subtle difference to the analogous representation of $TN$ in \citet{D24},
where $\Gamma$ is mentioned together with other components of $TN$ such as
${\mathcal M}, {\mathcal L}$ and ${\mathcal T}$ on the same level, see \citep[p.21]{D24}.
One could criticize that the question of whether symmetries leave the
gravitational constant invariant was thus decided from the outset.
In our approach it follows that $\Gamma$ will be uniquely determined:
For $N=2$ the r.~h.~s.~of
(\ref{eom2}) will be non-zero for all $i=1,2$ and $t\in(a,b)$. Hence the  l.~h.~s.~of
(\ref{eom2}) will also be non-zero and the positive scalar quantity $\Gamma$ is uniquely determined.
This mathematical result is the basis of the possibility to (approximately) measure $\Gamma$.

\subsection{Isomorphisms and automorphisms}\label{sec:NGIA}

We will not further specify the species of structure $\Sigma N$ belonging to the theory $TN$ in more detail,
because this can be easily reconstructed from the above information.
Similarly, we will only make conjectures about the automorphism group $\mbox{Aut}(A,B,\sigma)$
in  $\Sigma N$, although these are very likely. The usual calculations
showing that the Euclidean group generated by translations and rotations/reflections of ${\sf A}$
leaves the solution set of (\ref{eom2}) invariant can be employed to show that
the Euclidean group is a subgroup of $\mbox{Aut}(A,B,\sigma)$. Another subgroup
is the one-dimensional ``Euclidean group" generated by translations and reflections of the time axis $\sf{T}$.

\begin{figure}[ht!]
\centering
\includegraphics*[clip,width=0.7\columnwidth]{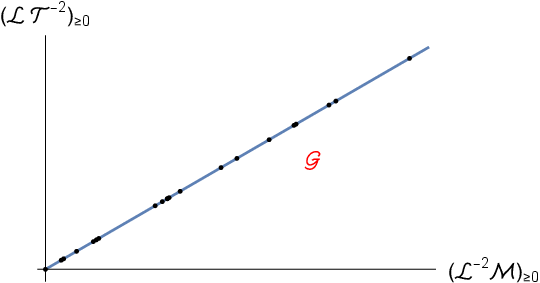}
\caption{Sketch of a structural element ${\mathcal G}$ of $TN$. The $x$-coordinate represents
the norm of the total gravitational force upon a particle divided by its mass
and by the gravitational constant $\Gamma$
(physical dimension $ \mbox{mass}\times \mbox{length}^{-2}$) denoted by $r_i(t)$ in (\ref{eomnorm}).
The $y$-coordinate represents the norm of its acceleration (physical dimension $ \mbox{length}\times \mbox{time}^{-2}$)
 denoted by $\ell_i(t)$ in (\ref{eomnorm}).
Possible gravitational experiments (some of which are indicated in this Figure)
give rise to points that lie on a ``universal line"  ${\mathcal G}$
with slope $\Gamma$. For special cases the total gravitational force upon a particle may vanish which is indicated
by the point with coordinates $(0,0)$.
}
\label{FIGG}
\end{figure}

However, with regard to the philosophical discussion about the status of certain physical quantities
mentioned in the Introduction, it is useful to closer analyze the three-dimensional
group of positive dilations $(\lambda, \tau, \mu)$ acting on ${\mathcal L}\times {\mathcal T}\times{\mathcal M}$
as isomorphisms or, possible automorphisms of structures $\langle A,B,\sigma\rangle$ of the species $\Sigma N$.

First consider the case of isomorphisms. The fundamental ranges ${\mathcal L}, {\mathcal T}$ and ${\mathcal M}$
are among the principal base sets $B_i,\,i=1,\ldots,b$ of $\Sigma N$ and without loss of generality we may set
$B_1={\mathcal L},\,B_2={\mathcal T},$ and $B_3={\mathcal M}$. Then we consider the  isomorphism
$f=(\mbox{id},\ldots,\mbox{id},f_1,\ldots,f_b)$ given by
$f_1=\lambda\cdot,\,  f_2=\tau\cdot$ and $f_3=\mu\cdot$ and $f_i=\mbox{id}$ for $i=4,\ldots,b$.
We have to calculate the transformed structural term $\sigma^f$ of $\Sigma N$.
One component of $\sigma$ is ${\mathbf X}$, a set of solutions of (\ref{eom2}). Since the fundamental
ranges do not occur in ${\mathbf X}$ we have ${\mathbf X}^f={\mathbf X}$. This is different
for the component $\sigma_m$ representing the mass function  $m:{\sf P} \to {\mathcal M}$.
Since $f_3=\mu\cdot$ operates on ${\mathcal M}$ and ${\sf P}$ is left unchanged we have in general $\sigma_m^f \neq \sigma_m$.
Other components of the structural term that are non-trivially transformed under $f$ are
the scalar products $\langle,\rangle_T: T\times T \to {\mathcal T}^2$ and
$\langle,\rangle_V: V\times V \to {\mathcal L}^2$ defined on the vector spaces $T$ and $V$ underlying
${\sf T}$ and ${\sf A}$. Hence it is clear that $\sigma^f\neq \sigma$ except for the special case $\lambda=\tau=\mu=1$
and $f$ will in general not be an automorphism of $\langle A,B,\sigma\rangle$. We then
consider the postulate of transportability of the axiom $\alpha(A,B,\sigma)$ of $\Sigma N$.
The interesting part of $\alpha(A,B,\sigma)$ will be the equality of the norm of both sides of (\ref{eom2}):
\begin{equation}\label{eomnorm}
\ell_i(t):= \| \ddot{\mathbf x}_i (t) \| = \Gamma  \left\|\sum_{j=1,j\neq i}^{N} m_j\;
  \frac{{\mathbf x}_j(t)-{\mathbf x}_i(t)}{\left|{\mathbf x}_j(t)-{\mathbf x}_i(t) \right|^3} \right\| =: \Gamma \,r_i(t)
  \;,
\end{equation}
The norm is defined as the square root of the scalar product of a vector with itself and hence
both sides of (\ref{eomnorm}) will be transformed by $f$. The $\ell_i(t)$ at the l.~h.~s.~will be multiplied by the numerical factor
$\lambda \tau^{-2}$ and the $r_i(t)$ at the r.~h.~s.~by the numerical factor $\mu\lambda^{-2}$.
Recall that in Axiom $TN$ \ref{AxTN}  $\Gamma$ was bounded by an existential quantifier.
Therefore (\ref{eomnorm}) is indeed equivalent to its transformed version and
$\Gamma$ is transformed by the factor $\lambda^3 \tau^{-2}\mu^{-1}$. This is
not in contradiction with the above result that the gravitational constant $\Gamma$
is uniquely determined in each model of $\Sigma N$ since the transformation $f$
maps each model of $\Sigma N$ onto an isomorphic, but different model.

Note also that the validity of the statement $\Gamma= 6.67430\times 10^{-11} m^3 s^{-2} kg^{-1}$
is not changed by the isomorphism $f$ since $f$ affects the value of $\Gamma$
as well as the units of length, time and mass.

Next we consider the case of an automorphism. We keep the notation of the last paragraphs
but denote the possible automorphism by
$g=(\mbox{id},\ldots,\mbox{id},g_1, g_2, \ldots, g_b)$
where $g_i=f_i$ for $i=1,2,3$.
The requirement $\sigma=\sigma^g$ enforces the action of $g_4: {\sf P}\to {\sf P}$ in the form
$g_4(p)=m^{-1}\left( \mu\, m(p)\right)$, where we have set $B_4:={\sf P}$.
Each particle is therefore mapped onto the particle with $\mu$ times the mass.
Similarly, the action of $g$ onto $B_5:={\sf A},\,B_6:= V$ and $B_7:={\sf T}, \, B_8:= T$
must be defined in such a way that the structural terms
$\langle,\rangle_V : V\times V \to {\mathcal L}^2$, $\langle,\rangle_T: T\times T \to {\mathcal T}^2$,
$\mbox{op}_V: V\times {\sf A}\to {\sf A}$ and $\mbox{op}_T: T\times {\sf T}\to {\sf T}$ remain invariant.
In particular, this means that we have to single out a point $A_0\in {\sf A}$ such that $g_5(A_0)=A_0$
and $g_5(A)-g_5(A_0)= g_6(A-A_0) = \lambda (A-A_0) $ for all $A\in{\sf A}$. Analogously for $g_7$ and $g_8$.
The choice of certain fixed points for spatial and temporal dilations
does not violate the translation invariance of the corresponding geometries,
since the set of possible automorphisms is formed by a union over all fixed points.

Then we consider the structural term ${\mathbf X}$ and split (\ref{eom2})
corresponding to some $(P,{\mathbf x})\in{\mathbf X}$ into two equations,
one for the directions and one for the magnitudes, the latter being (\ref{eomnorm}). The equation
for the directions is unchanged by $g$, except that the argument $t\in{\sf T}$ is replaced by
$g_7(t)$. Concerning the action of $g$ onto (\ref{eomnorm}) we again observe that the $\ell_i(t)$ at the l.~h.~s.~of
(\ref{eomnorm}) is multiplied by the factor $\lambda\,\tau^{-2}$,
whereas the $r_i(t)$ at the r.~h.~s.~are multiplied by the factor $\lambda^{-2}\,\mu$.
According to Axiom $TN$ \ref{AxTN} the points with these coordinates lie on a ``universal line" with slope
$\Gamma$, see Figure \ref{FIGG}. Invariance of this line under $g$ is satisfied iff both factors coincide, that is
\begin{equation}\label{lineinv}
\lambda\,\tau^{-2} = \lambda^{-2}\,\mu \quad \Leftrightarrow\quad \lambda^3\, \tau^{-2}\, \mu^{-1}=1
\;.
\end{equation}
This condition reduces the $3$-dimensional group of dilations to a two-dimensional subgroup
of ``similarity transformations" of the solutions of (\ref{eom2}).
If, for a given solution of (\ref{eom2}), we multiply all lengths by $\lambda>0$,
all time differences by $\tau>0$ and all masses by $\mu=\lambda^3\, \tau^{-2}$,
then we obtain another solution of (\ref{eom2}).
For $\mu=1$ and $N=2$ this special symmetry group implies in particular
Kepler's third law, see \citep[section 2.5]{A78}.
In any case, pure mass dilations, i.~e., dilations of the form $(1, 1, \mu)$
with $\mu\neq 1$ do not belong to the automorphism group because of the universality
of the gravitational constant $\Gamma$.

Having excluded Galilean transformations from the outset by our approach,
we conjecture that $\mbox{Aut}(A,B,\sigma)$ is generated by the above three types of subgroups,
the Euclidean groups acting on ${\sf A}$ and ${\sf T}$, and dilations satisfying (\ref{lineinv}).
We only mention in passing that symmetry groups play an important role in physics.
The symmetry groups mentioned above, which contain space- and time-translations as well as rotations,
are linked in classical mechanics via Noether's theorem with conservation laws
for momentum, energy and angular momentum, see \citep{A78}, and \citet{L71}.

\section{Active and passive interpretation of automorphisms}\label{sec:AP}

Although the term ``automorphism" of a structure of the species $\Sigma$ is initially of a purely meta-mathematical nature,
we have already alluded to its physical interpretation in the previous subsection \ref{sec:NGIA}.
In order to put this interpretation on a more solid basis, let us consider the reconstruction of applications of a physical
theory according to Ludwig (1990), albeit only sketchily and essentially limited to the example
of Newton's theory of gravitation according to Section \ref{sec:NGPP}.
The question arises as to the formal framework of these investigations.
While Ludwig prefers Bourbaki's original approach, as set out in \citet{B04},
and \citet{S79} is based on the set theory of Zermelo/Fraenkel,
I will limit myself to informal set theory in this paper.

Thus we assume that the base sets $A_i$ and $B_j$ as well as the components $\sigma_k$ of the structure term
$\sigma$ are already physically interpreted independently of the theory under consideration,
i.~e., in the terminology of Balzer, Moulines, and Sneed (1987), that they are ``non-theoretical terms".
For the example of Newtonian gravitation we may further restrict to only one auxiliary base set $A_1={\mathbb R}$.

Then, according to Ludwig, an application of the theory can be described by an
``observational report" (in the original: ``Realtext") ${\mathcal A}$.
The original intention when introducing ${\mathcal A}$ was to enable a statement
to be made as to whether ${\mathcal A}$ confirms the theory or not.
The naive criterion that the axiom of the theory, extended by ${\mathcal A}$, i.~e.,
$\alpha(A,B,\sigma)\wedge {\mathcal A}$,
remains consistent is too simple because it does not take experimental and other inaccuracies into account.
For the purposes of this paper, however, we can leave aside the problem of a meta-theory of confirmation.
Instead, we will use the naive criterion of confirmation, fully aware of its inadequacy,
to simplify the explanation of the physical interpretation of automorphisms.

${\mathcal A}$ consists of a conjunction of atomic sentences of the following form. On the one hand
different types of physical objects are typified by ``$x_{kj}\in A_j$" and ``$\lambda_\ell\in {\mathbb R}$".
The $\lambda_\ell$ are not variables but concrete numbers such as $\lambda_1=-2.8$.
Since we only have one auxiliary base set $A_1={\mathbb R}$, the statements
``$\lambda_\ell\in {\mathbb R}$" are actually superfluous.

In addition, there are statements that represent the results of measurements.
They have the form ``$X_{\nu s}\in \sigma_s$", where the $X_{\nu s}$ are composed
from the $x_{kj}$ and the $\lambda_\ell$ according to the echelon construction schemes
$\tau_s$ with $\sigma_s\in \tau_s({\mathbb R},B)$.
As an example, let $\sigma_1\in{\mathcal P}({\sf P}\times {\mathcal M})$
be the structural component corresponding to the mass function $m:{\sf P}\longrightarrow{\mathcal M}$,
and further let the typifications ``$p\in{\sf P}$" and ``$\mu\in{\mathcal M}$" be part of ${\mathcal A}$.
Then the measurement result that the point particle $p$ has the mass $\mu$
is expressed by ``$(p,\mu)\in\sigma_1$".
Remember that $\mu$ is not a number but the value of a quantity of dimension ``mass".
Hence it is sensible that there will be another sentence in ${\mathcal A}$
of the form, say, ``$(2.3, \mu,\mu_1) \in \sigma_2$".
Here $\sigma_2\in{\mathcal P}\left({\mathbb R}\times {\mathcal M}\times {\mathcal M}\right)$
is the structural component corresponding to the external multiplication
${\mathbb R}\times {\mathcal M}\longrightarrow{\mathcal M} $ in the one-dimensional vector space ${\mathcal M}$
and $\mu_1\in {\mathcal M}$ denotes a special mass that is used as a mass unit.
Taken together, the sentences considered so far in ${\mathcal A}$ state that the point mass
$p$ has the mass $\mu=2.3\,\mu_1$.

Of course, one cannot infer from these statements that $\mu_1\in{\mathcal M}$ is intended as a unit of mass.
We therefore propose, going beyond Ludwig, that the atomic statements in ${\mathcal A}$ can be decomposed
into two parts ${\mathcal F}$ and ${\mathcal B}$ such that
\begin{equation}\label{AFB}
 {\mathcal A} = {\mathcal F} \wedge {\mathcal B}
 \;.
\end{equation}
Here, ${\mathcal F}$ (the ``frame") contains all sentences that are not real measurement results
but serve to prepare such measurement results. In the case of Newtonian gravitation ${\mathcal F}$
contains, besides the mentioned sentence ``$\mu_1\in {\mathcal M}$" analogous sentences concerning
the units of length and time as well as sentences denoting coordinate systems for space and time.
The spatial coordinate system is given, for example, by four points $e_i\in A,\, i=0,1,2,3$,
which span an orthonormal affine basis in $A$ (``normal" w.r.~t.~a distinguished unit of length).
Those measurement results that guarantee the latter are also part of ${\mathcal F}$.
Since we do not want to develop a general meta-theory in which
${\mathcal F}$ is characterized more precisely, we must limit ourselves to these exemplary remarks.

In any case, the division of ${\mathcal A}={\mathcal F}\wedge {\mathcal B}$ has the consequence
that the statements of ${\mathcal B}$ (the ``proper observational report")
can be interpreted completely numerically. If coordinate systems
and units are defined, the measurements can be read in such a way that the $p$th point particle at different times
$t_\nu\in{\mathbb R}$ has the positions ${\mathbf x}_p(t_\nu)\in {\mathbb R}^3$.

Now we consider an automorphism $f:({\mathbb R},B)\longrightarrow({\mathbb R},B)$.
Recall that $f$ is a shorthand for the $b+1$-tuple of bijections
$f=\left\langle \mbox{id}, f_1,\ldots, f_b\right\rangle,\; f_j:B_j\longrightarrow B_j$.
There is some obvious sense in which $f$ operates on ${\mathcal A}$:
Whenever the denotation $x_{kj}$ of a physical object occurs in ${\mathcal A}$
it will be replaced by $f_j(x_{kj})$, whereas the numbers $\lambda_\ell$ occurring in ${\mathcal A}$
are left unchanged. Let the observational report transformed in this way  be denoted by ${\mathcal A}^f$,
then it follows that $\alpha({\mathbb R},B,\sigma)\wedge {\mathcal A}$ is consistent iff
$\alpha({\mathbb R},B,\sigma)\wedge {\mathcal A}^f$ is consistent.  The proof only uses
the transportability of $\alpha({\mathbb R},B,\sigma)$ as defined in Section \ref{sec:SS}
and not that $f$ is an automorphism of $({\mathbb R},B)$.
The transformed observational report ${\mathcal A}^f$ is essentially the same as ${\mathcal A}$
with the only difference that the denotations $x_{kj}$ are replaced by $x_{kj}'=f_j(x_{kj})$.
We will call this interpretation
of $f$ the ``Leibniz interpretation" since we think that it comes close to what Leibniz
had in mind in his argument against absolute space: All physical objects are transformed in a certain
way while the principal base sets are left fixed without changing the affair of states.
The Leibniz interpretation could also applied to isomorphisms $f:({\mathbb R},B)\longrightarrow({\mathbb R},B)$
that are not automorphisms.

I think, however, that other physical interpretations of the automorphisms $f$ are possible and important.
For a fixed frame ${\mathcal F}$ it is clear how $f$ operates on the
numerical positions given by points in ${\mathbb R}^3$, as well as on the times
and ranges of physical quantities, which are represented by different copies of ${\mathbb R}$.
This also allows one to define an effect of $f$ on the proper measurements
that are summarized in the statement ${\mathcal B}$.
For example, if $f$ involves a doubling of the masses (accompanied by suitable dilations of length and time),
then the above-mentioned sentence
``$(2.3, \mu,\mu_1) \in \sigma_2$" will be mapped by $f$ onto the sentence
``$(4.6, f(\mu),\mu_1) \in \sigma_2$". Note that
the unit of mass $\mu_1$, which was introduced in ${\mathcal F}$, remains unchanged.
Analogously, all denotations $x_{kj}$ of a physical object occurring in ${\mathcal B}$, but not in ${\mathcal F}$,
are replaced by $f_j(x_{kj})$ whereas the numbers $\lambda_\ell$ occurring in ${\mathcal B}$
are replaced by some numbers $\lambda'_\ell$ to be calculated by the effect of $f$ on the numerical
measurement results. This defines a transformed proper observational report ${\mathcal B}'$
which, together with the unchanged frame ${\mathcal F}$, yields ${\mathcal A}'={\mathcal F}\wedge {\mathcal B}'$.
Again adopting the naive confirmation criterion, one can show that $\alpha({\mathbb R},B,\sigma)\wedge {\mathcal A}$
is consistent iff $\alpha({\mathbb R},B,\sigma)\wedge {\mathcal A}'$
is so. This time the proof crucially uses the condition that $f$ is an automorphism of $({\mathbb R},B)$.
If ${\mathcal A}$ was the report of an actual measurement then ${\mathcal A}'$ would be the report of
another fictitious measurement. Sometimes ${\mathcal A}'$ can be realized, sometimes not, for example if
$f$ involves a time translation into the past. But there are no physical reasons why ${\mathcal A}'$ should
not be realizable. In this sense $f$ maps possible experiments onto possible experiments.
Of course, the latter statement ignores the possible limits of the theory under consideration.
If a variant of the present formulation of Newtonian gravitation allows boosts as automorphism, then
boosts to velocities exceeding the velocity of light are physically impossible.

The difference to the above-sketched Leibniz interpretation of an automorphism $f$ is that
this time the frame containing units and coordinate systems is left fixed while the remaining physical
objects and numbers are transformed. We will call this the ``active interpretation" of an automorphism 
and symbolize it by
\begin{equation}\label{symbolactive}
 {\mathcal A}= {\mathcal F}\wedge  {\mathcal B} \stackrel{f,\mbox{ active}}{\xrightarrow{\hspace*{12mm}}}
 {\mathcal A}'= {\mathcal F}\wedge  {\mathcal B}'
 \;.
\end{equation}

The wording suggests that there is also a passive interpretation of an automorphism $f$.
It will be convenient to rather describe the passive interpretation of $f^{-1}$ given the active
interpretation of $f$. To this end we again consider the actively transformed observational report
$ {\mathcal A}'= {\mathcal F}\wedge  {\mathcal B}'$ and make the following changes:
Each denotation of a ``frame object", i.~e.~of some $x_{kj}$ occurring in ${\mathcal F}$ is replaced by $f_j^{-1}(x_{kj})$
and likewise for all occurrences of this $x_{kj}$ in  ${\mathcal B}'$. In contrast, all remaining denotations
of objects occurring in  ${\mathcal B}'$ and all numbers $\lambda_\ell$ occurring in ${\mathcal A}'$ are left unchanged.
This defines the ``passively transformed observational report" ${\mathcal A}''={\mathcal F}''\wedge  {\mathcal B}''$.

Consider the above example of  ``$(2.3, \mu,\mu_1) \in \sigma_2$" in ${\mathcal A}$
and the actively transformed statement ``$(4.6, f(\mu),\mu_1) \in \sigma_2$" in ${\mathcal A}'$.
The corresponding passively transformed statement in ${\mathcal A}''$
would be ``$(4.6, \mu,f^{-1}(\mu_1) \in \sigma_2$". This means that,
compared with ``$(2.3, \mu,\mu_1) \in \sigma_2$" ,
 the value of the mass quantity $\mu$ is unchanged
but the unit of mass $\mu_1$ is halved. Therefore the numerically value of $\mu$ and hence of all point masses
occurring only in ${\mathcal B}$ are again doubled. We will call this the ``passive interpretation" of an automorphism 
and symbolize it by
\begin{equation}\label{symbolactive}
 {\mathcal A}= {\mathcal F}\wedge  {\mathcal B} \stackrel{f,\mbox{ passive}}{\xrightarrow{\hspace*{12mm}}}
 {\mathcal A}''= {\mathcal F}''\wedge  {\mathcal B}''
 \;.
\end{equation}

We hope that our definitions capture the notions of ``active transformations'' and ``passive transformations''
informally used in physics. As indicated in the Introduction, we also believe that the debate
on the metaphysical status of quantities using ``symmetry arguments'' could benefit
from the above distinction between Leibniz, active and passive interpretations of automorphisms.

Although, as we said, we do not want to enter into the discussion about the
metaphysical status of physical quantities, we would like to venture
a guess here as to how the above reconstruction could contribute to clarifying this discussion.
Consider a constant dilation of all masses with the factor $\mu>0$,
without further accompanying transformations, i.e. {\it ceteris paribus}.
According to our approach, there seems to be only one way to construe this dilation in the theory $TN$.
The exclusive transformation $f_3:{\mathcal M}\to {\mathcal M},\;M\mapsto \mu\,M$ must be understood
as an isomorphism and interpreted in the above sense of Leibniz. Thus the physics is
unchanged, in particular the escape velocity of a body on the earth's surface remains constant.
The latter follows since the value of gravitational constant $\Gamma$ is forced to transform with the
factor $\mu^{-1}$, as it follows from the discussion in subsection \ref{sec:NGIA}.
We emphasize again that the statement $\Gamma= 6.67430\times 10^{-11} m^3 s^{-2} kg^{-1}$
is left unchanged and that a body of mass $1 kg$  has still the mass $1 kg$ after the dilation,
since the value of the body's mass, which represents the mass unit ``kg", is also transformed.

It may be that someone is not satisfied with this interpretation and insists that the mass scaling should
transform a body of $1\,kg$ into a body of $\mu\, kg$. The formulation of \citep[p.9]{M24} who
speaks of ``(Exclusive) active Leibniz mass scaling" could be understood in this sense.
According to our approach this would mean the attempt to view the exclusive mass dilation as an
automorphism in the theory $TN$ and to interpret it in the active sense described above.
Then the answer would be that this is not possible, see again the discussion in subsection \ref{sec:NGIA}.
In this case the argument with the escape velocity of a body applies.
Ironically, the reason this time lies in the necessary invariance of $\Gamma$,
because $\Gamma$ is a theoretical term in $TN$ that proves to be uniquely determined
within a model $\langle A,B,\sigma\rangle$ of $\Sigma N$.

\section{Summary and Outlook}\label{sec:SO}
This work is based on the observation that structuralist reconstructions
have so far failed to capture an essential aspect of physics, namely the structure of
physical quantities with a dimension and the corresponding dimensional calculus.
On the one hand, the consistent account of dimensional quantities is crucial
for the understanding of reduction relations between theories, which are symbolically expressed
by limits as  ``$\hbar \to 0$" or ``$c\to\infty$", see \citet{S98}.
Moreover, these structures are also relevant for recent debates on the metaphysics of quantities.
On the other hand, there is no reason why the dimensional calculus
cannot be incorporated into the species of structure concept used to reconstruct physical theories.
This is demonstrated by two examples, a toy example of an ``Ohmian theory",
and a specialization of Newtonian mechanics to a system of $N$ gravitating point masses.
Following Tao (2013) we propose to represent dimensional quantities
by one-dimensional, real, ordered vector spaces ${\mathcal R}$ so that
the choice of a physical unit corresponds to the choice of a positive basis in ${\mathcal R}$.
The dimensional calculus including the $\Pi$-theorem can then be reconstructed using tensor products and dual spaces.

The objection could be raised against our reconstruction that it sets up a
formal apparatus in order to provide an intricate explanation of a practice
that is so simple that it is already mastered by pupils in physics lessons.
Anyone who has understood Ohm's law might doubt it again with our
reconstruction in  Section \ref{sec:Ohm}.
Of course, we do not want to recommend this reconstruction for physics lessons.
Nor do we want to claim that physicists do not know what they are doing
when they calculate with physical units. Our concern is rather the following:
If philosophers (or physicists) propose a structuralist reconstruction of
physical practice that includes dealing with mathematics,
then this should consequently include dealing with dimensional quantities.

The benefit of our proposal for physics could be that it provides a
systematic foundation for physical reasoning usually based on intuition and ``hand waving".
Furthermore, we believe that the recent philosophical debate about the
metaphysical status of status of quantities could benefit from a clarification of the basic concepts,
as proposed here. We have given an example of this at the end of Section \ref{sec:AP}.

We concede that the theory of physical dimensions in the formalism of species of structure
would have to be further developed.
We have proposed a concept for $3$-dimensional physical quantities
and a mathematical reconstruction of the formation of rational powers of physical quantities,
but we have not yet dealt with relativistic theories and tensors with physical dimension.
Similarly, we have not discussed how to evaluate the elimination of natural constants
by the choice of ``natural units" (e.g. Planck units).
The section on the physical interpretation of ``automorphisms'' had to remain sketchy.
So there is still some work to be done.

\section*{Acknowledgment}\label{sec:ACK}

I would like to thank Felix M\"uhlh\"olzer, Simon Friederich and
Caspar Jacobs for critically reviewing earlier versions of this paper
and for references to the relevant philosophical literature,
Daniel Burgarth and David Gross for discussions on the topic
of physical dimensions, and the latter especially for the reference to \citet{T13}.

\begin{appendices}

\section{Proof of Theorem \ref{TPi}}\label{sec:Pi}

The vectors in the kernel $\ker(A)$ of the integer-valued matrix $A$ can be obtained
by solving a linear, homogeneous system of equations, e.~g., by Gaussian elimination.
This shows that the kernel of $A$ has a basis of vectors with rational coefficients,
hence, after multiplication by the least common multiple of the denominators, with integer coefficients.
Such a basis of $\ker(A)$ will be denoted by $\left( {\mathbf p}^{(k)}\right)_{k=1,\ldots,K}$.
The coefficients of the basis vectors $p^{(k)}_\ell\in{\mathbb Z}$ form an $L\times K$-matrix.
Let $R$ denote the rank of $A$, i.~e.~the dimension of its image $\mbox{im}(A)\subset {\mathbb R}^N$,
which equals the rank of $A^\top$,
By the rank-nullity theorem, see \citet{S93}, we have $K+R=L$.

By inserting the above-defined coefficients  $p^{(k)}_\ell$ into (\ref{Pigroup}) we obtain
the ``$\Pi$-group" of quantities with values $\left(\Pi_1,\ldots, \Pi_K\right)$. They are
dimensionless due to
\begin{equation}\label{Pidimless}
 \Pi_k= \prod_{\ell=1}^{L}Q_\ell^{ p_{\ell}^{(k)}}=
 \prod_{\ell=1}^{L}\left(\prod_{i=1}^N R_i^{A_{i \ell}} \right)^{ p_{\ell}^{(k)}}
 = \prod_{i=1}^N R_i^{\sum_\ell A_{i \ell} p_{\ell}^{(k)}}
\end{equation}
and $\sum_\ell A_{i \ell} p_{\ell}^{(k)}=0$ since ${\mathbf p}^{(k)}\in \ker A$ for all $k=1,\ldots,K$.

Next we translate the condition that the physical law of the form
$f(Q_1,\ldots, Q_L)=0$ is dilationally invariant into the language of affine geometry.
Two points $A,B\in {\mathcal Q}_+$ are connected by a dilation of the form (\ref{actionQ})
iff $A\sim B$ where the equivalence relation is defined by the subspace
${\mathcal U}=\mbox{im} A^\top\subset {\mathbb R}^L$ in the sense of
$A\sim B\Leftrightarrow \stackrel{\xrightarrow{\hspace*{0.5cm}}}{A,B}\in {\mathcal U}$.
The physical law defines a subset
\begin{equation}\label{sublaw}
 {\mathcal S}=\{(Q_1,\ldots, Q_L)\in \widetilde{\mathcal Q}\left|\right. f(Q_1,\ldots, Q_L)=0\}
 \subset \widetilde{\mathcal Q}  \subset  {\mathcal Q}_+
 \;.
\end{equation}
Its dilational invariance can be written as $A\in {\mathcal S}\Leftrightarrow [A]_\sim \subset {\mathcal S}$.
Therefore a dilationally invariant law can also be represented by a subset of equivalence classes
\begin{equation}\label{sublaw}
 {\mathcal S}'=\{[A]_\sim \in  {\mathcal Q}_+/_\sim\left| [A]_\sim \subset {\mathcal S}\right.\}
 \;.
\end{equation}

The definition of the $\Pi$-group (\ref{Pigroup}) induces an affine map $\psi: {\mathcal Q}_+\to {\Pi}_+$,
where $\Pi_+=\prod_{k=1}^{K}{\mathbb R}_{>0}$ is viewed as an affine space w.~r.~t.~the action
$(\Pi_1,\ldots,\Pi_K)\mapsto (e^{x_1}\Pi_1,\ldots,e^{x_K}\Pi_K)$  of ${\mathbb R}^K$.\\

\begin{lemma}\label{Lpsi}
\begin{enumerate}
  \item $\psi: {\mathcal Q}_+\to {\Pi}_+$ is  constant on $\sim$-equivalence classes
  and hence  can be factorized as $\psi=\chi\circ \pi$ where
  $ {\mathcal Q}_+\stackrel{\pi}{\longrightarrow}{\mathcal Q}_+/_\sim\stackrel{\chi}{\longrightarrow}\Pi_+$.
  \item The affine map $\chi:{\mathcal Q}_+/_\sim\to \Pi_+$ is bijective.
\end{enumerate}
\end{lemma}

\noindent {\bf Proof}:
\begin{enumerate}
  \item Let $A\sim A'$, that is,
  $A=\left( Q_1,\ldots,Q_L\right) \in {\mathcal Q}_+$ and
  $A'=\left( \Lambda_1 Q_1,\ldots,\Lambda_L Q_L\right) \in {\mathcal Q}_+$
  where $\Lambda_\ell=\prod_i \lambda_i^{A_{i\ell}}$ for $\ell=1,\ldots, L$. Moreover, let
  $\psi(A)=\left( \Pi_1,\ldots,\Pi_K\right)$ and $\psi(A')=\left( \Pi'_1,\ldots,\Pi'_K\right)$.
  Then it follows that for all $k=1,\ldots,K$:
  \begin{equation}\label{proof1}
   \Pi'_k = \prod_{\ell} \left( \Lambda_\ell Q_\ell\right)^{p_\ell^{(k)}}=
   \prod_{\ell i} \left( \lambda_i^{A_{i\ell}}\right)^{p_\ell^{(k)}}\,\prod_\ell Q_\ell ^{p_\ell^{(k)}}
   =\left(\prod_i \lambda_i^{\sum_\ell A_{i\ell}p_\ell^{(k)}}\right)\,\Pi_k =\Pi_k
   \;,
  \end{equation}
  since $\sum_\ell A_{i\ell}\,p_\ell^{(k)}=0$ for $k=1,\ldots,K$. Hence $\psi$ is constant on $\sim$-equivalence classes.
  \item We will first show that $\chi$ is injective. This is equivalent to $\psi(A)=\psi(A')\Rightarrow A\sim A'$.
  Let us assume  $\psi(A)=\psi(A')$. With the same notation as in the first item,
  but setting $A'=\left(Q_1',\ldots,Q_L'\right)=\left(e^{y_1}\,Q_1,\ldots, e^{y_L}\,Q_L\right)$,
  we conclude that for all $k=1, \ldots,K$,
  \begin{eqnarray}\label{proof2a}
    \Pi'_k &=&\prod_{\ell} \left( Q'_\ell\right)^{p_\ell^{(k)}}=
    \prod_{\ell} \left( e^{y_\ell}\,Q_\ell\right)^{p_\ell^{(k)}}=\left(e^{\sum_\ell y_\ell p_\ell^{(k)}}\right) \Pi_k =\Pi_k\\
    \label{proof2b}
    &\Leftrightarrow& \sum_\ell y_\ell p_\ell^{(k)}=0 \mbox{ for all } k=1,\ldots,K
    \Leftrightarrow {\mathbf y} \perp \ker A  \Leftrightarrow  {\mathbf y}\in \mbox{im} \left(A^\top \right)\\
    \label{proof2b}
    & \Rightarrow& A\sim A'
     \;,
  \end{eqnarray}
  where we have used the identity $ \left(\ker A\right)^\perp = \mbox{im} \left(A^\top \right)$, see \citet{S93}.
  This proves that $\psi$ is injective. Moreover, since $\dim {\mathcal Q}_+/_\sim\ =\dim \Pi_+ = K$,
  $\chi$ is also bijective.
\end{enumerate}
This completes the proof of Lemma \ref{Lpsi}. \hfill$\Box$\\

After having shown that ${\mathcal Q}_+/_\sim$ and $\Pi_+$ are isomorphic affine spaces,
we can just as well represent the physical law under consideration by the subset
\begin{equation}\label{defSpp}
 {\mathcal S}'':= \chi\left[ {\mathcal S}'\right]\subset \Pi_+
 \;,
\end{equation}
such that the following equivalences hold for all $A\in {\mathcal Q}_+$:
\begin{equation}\label{equivS}
  A\in {\mathcal S}  \Leftrightarrow    [A]_\sim \in {\mathcal S}'
  \Leftrightarrow  \chi\left([A]_\sim \right))\in {\mathcal S}''
  \Leftrightarrow  (\chi\circ\pi)(A)\in {\mathcal S}''
 \Leftrightarrow  \psi(A)\in {\mathcal S}''
 \;.
\end{equation}
Finally, we choose a function $F:\Pi_+\to {\mathbb R}$ such that
\begin{equation}\label{FSpp}
 F(\Pi_1,\ldots,\Pi_K) =0   \Leftrightarrow (\Pi_1,\ldots,\Pi_K)\in  {\mathcal S}''
 \;,
\end{equation}
which completes the proof of Theorem \ref{TPi}.

\section{Axiomatic characterization of positive ranges}\label{sec:ACP}

For some questions raised in this work it is useful to describe the transition from positive ranges
to (signed) ranges of physical quantities more precisely.
This topic is usually dealt with in the ``Theory of measurement", see for example  \citet{KLST71};
we formulate a shortened version of it here, which is suitable for our purposes.

A ``positive range" $\left({\sf P},+,\cdot\right)$ is a set ${\sf P}$
with a binary operation $p+q$ and an external multiplication $\lambda \cdot p$
for $p,q\in {\sf P}$ and $\lambda\in {\mathbb R}_{>0}$ satisfying
\begin{eqnarray}
\label{ax1}
  p+q &=& q+p, \\
  \label{ax2}
  (p+q)+r &=& p+(q+r), \\
  \label{ax3}
  \lambda\cdot(\mu\cdot p)&=& (\lambda \mu)\cdot p,\\
  \label{ax4}
  \lambda\cdot p= p &\Leftrightarrow& \lambda=1,\\
  \label{ax5}
  \lambda\cdot(p+q)&=& \lambda\cdot p + \lambda\cdot q,\\
  \label{ax6}
  (\lambda+\mu)\cdot p &=& \lambda\cdot p + \mu\cdot q
  \;.
\end{eqnarray}
The one-dimensionality of ${\sf P}$ is captured by the additional axiom
\begin{equation}\label{ax7}
  \mbox{For all } p,q\in {\sf P}\mbox{ there exists some } \lambda\in {\mathbb R}_{>0}
  \mbox{ such that } p=\lambda\cdot q
  \;.
\end{equation}
The positive range  ${\sf P}$ has a total order defined by
$p<q\,\stackrel{\rm def}{\Leftrightarrow} \,p=\lambda\cdot q$ and $0<\lambda <1$, or, equivalently, by the condition
that there exists some $r\in {\sf P}$ such that $q=p+r$.
A couple of consequences from these axioms can be derived. As an example we
prove the so-called shortening rule:
For all $p_1,p_2,q\in {\sf P}$ there holds
\begin{equation}\label{shortrule}
 \mbox{If } p_1+q=p_2+q\quad \mbox{then} \quad p_1=p_2
 \;.
\end{equation}
This follows from
\begin{eqnarray}
 && p_1+q = p_2+q \\
  &\stackrel{(\ref{ax7})}{\Rightarrow}& \lambda_1\cdot q+q=\lambda_2\cdot q+q \\
   &\stackrel{(\ref{ax4}),(\ref{ax6})}{\Rightarrow}& (\lambda_1+1)\cdot q=(\lambda_2+1)\cdot q \\
    &\stackrel{(\ref{ax3})}{\Rightarrow}&\frac{\lambda_1+1}{\lambda_2+1} \cdot q= q \\
     &\stackrel{(\ref{ax4})}{\Rightarrow}&\frac{\lambda_1+1}{\lambda_2+1}=1\\
     &\Rightarrow& \lambda_1=\lambda_2\\
      &\Rightarrow& p_1=p_2
      \;.
\end{eqnarray}

We want to embed a positive range ${\sf P}$ into a (signed) range of a physical quantity ${\mathcal R}$
which has been defined previously as a one-dimensional real ordered vector space.
Intuitively, we consider the pairs $(p,q)\in {\sf P}\times  {\sf P} $  as differences
corresponding to elements of ${\mathcal R}$. However, different
pairs will correspond to the same element of ${\mathcal R}$.
Hence we define an equivalence relation $\sim$ on ${\sf P}\times  {\sf P} $ by
\begin{equation}\label{defsim}
  (p_1,q_1) \sim (p_2,q_2)  \stackrel{\rm def}{\Leftrightarrow} p_1+q_2 = q_1 + p_2
  \;.
\end{equation}
and define ${\mathcal R}$ as the corresponding set of equivalence classes:
\begin{equation}\label{defR}
 {\mathcal R}:= \left( {\sf P}\times  {\sf P} \right)_\sim
 \;,
\end{equation}
The elements of  ${\mathcal R}$ will be denoted by $R=[p,q]$, where the pair
$(p,q)$ is any representative of the corresponding equivalence class $R$.

Next we will define an addition $R_1+R_2$ and an external multiplication
$\lambda\cdot R, \; \lambda\in {\mathbb R}$ on ${\mathcal R}$.
This can be accomplished by
\begin{equation}\label{defplusR}
 [p_1,q_1]+[p_2,q_2]:= [p_1+p_2,q_1+q_2]
 \;,
\end{equation}
and
\begin{equation}\label{defmultR}
\lambda\cdot[p,q]:=\left\{
\begin{array}{r@{\quad : \quad}l}
   \left[\lambda\cdot p,\lambda\cdot q\right] & \lambda >0 \\
   \left[|\lambda|\cdot q,|\lambda|\cdot p\right]& \lambda <0\\
   \left[p,p\right]& \lambda=0\;.
 \end{array}
 \right.
 \end{equation}
Moreover, we define an order $<$ on ${\mathcal R}$ by
\begin{equation}\label{deforder}
  [p_1,q_1]< [p_2,q_2] \stackrel{\rm def}{\Leftrightarrow} p_1+q_2 < q_1+p_2
  \;.
\end{equation}
Of course, it can be shown that  these definitions are independent
on the chosen representatives in $[p,q]$.

W.~r.~t.~these definitions one can show that $\left({\mathcal R},+,\cdot\right)$ fulfils the axioms
of a one-dimensional real ordered vector space. We will only mention that the
zero element of ${\mathcal R}$ w.~r.~t.~addition is given by $0=[p,p]$, and hence
the negative of $[p,q]$ will be $[q,p]$ which explains the definitions in (\ref{defmultR}).
The positive part of ${\mathcal R}$ is, as usual, denoted by ${\mathcal R}_{>0}$.

One can embed ${\sf P}$ into the positive part of  ${\mathcal R}$.
More precisely, there exists a ``canonical" bijection $\jmath: {\sf P} \rightarrow {\mathcal R}_{>0}$
which is an isomorphism of positive ranges w.~r.~t.~the addition $+$ and the external multiplication $\cdot$.
It can be defined by
\begin{equation}\label{defembed}
  \jmath(p):= [q+p,q],\quad \mbox{for all}\quad p\in{\sf P}
  \;,
\end{equation}
which is independent of the choice of $q\in{\sf P}$. We will only sketch the
proof that $\jmath$ is injective. Thus assume $[q+p_1,q]=[q+p_2,q]$ for some
$p_1,p_2,q\in{\sf P}$. Then we have
\begin{eqnarray}
  [q+p_1,q]=[q+p_2,q]
   &\stackrel{(\ref{defsim})}{\Rightarrow}& (q+p_1)+q=q+(q+p_2) \\
    &\stackrel{(\ref{ax1}),(\ref{ax4}),(\ref{ax6})}{\Rightarrow}& 2\cdot q+p_1=2\cdot q+p_2 \\
     &\stackrel{(\ref{shortrule})}{\Rightarrow}& p_1=p_2
     \;.
\end{eqnarray}
Although we have not proven all details we may summarize the foregoing considerations in the following:\\

\begin{prop}\label{P1}
For every positive range, i.~e., for every triple $\left({\sf P},+,\cdot\right)$ satisfying the axioms
(\ref{ax1}) - (\ref{ax7}), there exists a one-dimensional real ordered vector space ${\mathcal R}$
and a canonical isomorphism $\jmath: {\sf P} \rightarrow {\mathcal R}_{>0}$.
\end{prop}

\section{Rational powers of physical quantities }\label{sec:RE}

For this Section we consider a fixed range of a physical quantity ${\mathcal R}$, fundamental or derived,
and a positive rational number $q=\frac{m}{n}$, such that $m$ and $n$ are relatively prime positive integers.
The aim is to define a new physical quantity ${\mathcal Q}={\mathcal R}^q$,
the $q$-th power of ${\mathcal R}$.

To this end we consider the set of homogeneous functions of degree $q$
\begin{equation}\label{defMq}
 {\mathcal H}^q := \{f:{\mathcal R}_{>0} \rightarrow  {\mathbb R}_{>0} \left| \right. f(\kappa r)=\kappa^q\,f(r)\quad\mbox{for all  }
 r\in {\mathcal R}_{>0} \mbox{ and  } \kappa>0\}
 \;.
\end{equation}
${\mathcal H}^q$ can be equipped with an addition $f_1+f_2$
and an external multiplication  $f\mapsto \lambda f, \,\lambda>0$
in an obvious way and will form a positive range w.~r.~t.~these operations in the sense of
Appendix \ref{sec:ACP}.
Hence ${\mathcal H}^q$ can be identified with the positive part
${\mathcal S}^q_{>0}$ of a range of a physical quantity ${\mathcal S}^q$, see Proposition \ref{P1}.
Then we define ${\mathcal Q}={\mathcal R}^q$ as the dual space of ${\mathcal S}^q$:
\begin{equation}\label{defQ}
 {\mathcal Q}:={\mathcal R}^q :=\left( {\mathcal S}^q\right)^\ast\;.
\end{equation}
Hence also  ${\mathcal Q}$ is the range of a physical quantity with positive part ${\mathcal Q}_{>0}$.

It remains to show that the above definition is adequate.
For this it would suffice to show that there is a canonical isomorphism
${\mathcal R}^m \cong {\mathcal Q}^n $ or, equivalently,
$\left({\mathcal R}^m\right)^\ast \cong \left({\mathcal Q}^n\right)^\ast$.

Consider an arbitrary unit $R_0\in {\mathcal R}_{>0}$.
To this we can assign a unit $\hat{R}_0\in {\mathcal Q}_{>0}$
by the definition
\begin{equation}\label{defRhat}
 \hat{R}_0(f):= f(R_0)>0\quad \mbox{for all  } f\in {\mathcal S}^q_{>0}=  {\mathcal H}^q
 \;,
\end{equation}
where we have used that it suffices to define the action of $\hat{R}_0$
for the positive part of ${\mathcal S}^q$. Then we consider the $1:1$ relation
between elements $f\in \left({\mathcal R}^m\right)^\ast$  and elements
$g\in \left({\mathcal Q}^n\right)^\ast$ given by the equation
\begin{equation}\label{canonicaliso}
 f(\underbrace{R_0\otimes\ldots\otimes R_0}_{m\, \mbox{\scriptsize factors}})=
 g(\underbrace{\hat{R}_0\otimes\ldots\otimes \hat{R}_0}_{n\, \mbox{\scriptsize factors}})
 \;.
\end{equation}
Clearly, (\ref{canonicaliso}) defines a linear order isomorphism between
the one-dimensional real ordered vector spaces
$\left({\mathcal R}^m\right)^\ast$ and $\left({\mathcal Q}^n\right)^\ast$.
It remains to show that this isomorphism is ``canonical", which here means
that its definition does not depend on the choice of the unit $R_0$.

Hence let $R_1\in {\mathcal R}_{>0}$ be another unit, necessarily satisfying
$R_1=\kappa\,R_0$ for some $\kappa>0$.
Let $f\in  {\mathcal H}^q $ be arbitrary, then it follows that
\begin{equation}\label{R1R0}
 \hat{R}_1(f)=f(R_1)=f(\kappa\,R_0) = \kappa^q\,f(R_0) =\kappa^q\,\hat{R}_0(f)
 \;,
\end{equation}
and hence
\begin{equation}\label{hatR1}
  \hat{R}_1 = \kappa^q\,\hat{R}_0 ,\quad \mbox{and}\quad \hat{R}_0 = \kappa^{-q}\,\hat{R}_1
  \;.
\end{equation}
Then the claim follows from
\begin{eqnarray}
  f\left(R_1\otimes\ldots\otimes R_1 \right) &=&  f\left(\kappa R_0\otimes\ldots\otimes \kappa R_0 \right) \\
  &=& \kappa^m\, f\left(R_0\otimes\ldots\otimes R_0 \right) \\
  &\stackrel{(\ref{canonicaliso})}{=}& \kappa^m\, g\left(\hat{R}_0\otimes\ldots\otimes\hat{R}_0 \right) \\
   &\stackrel{(\ref{hatR1})}{=}& \kappa^m\, g\left(\kappa^{-q}\hat{R}_1\otimes\ldots\otimes\kappa^{-q}\hat{R}_1 \right) \\
   &=& \underbrace{\kappa^m\,\kappa^{-n q}}_{\kappa^m\,\kappa^{-m}=1}g\left(\hat{R}_1\otimes\ldots\otimes\hat{R}_1 \right) \\
   &=& g\left(\hat{R}_1\otimes\ldots\otimes\hat{R}_1 \right)
   \;.
\end{eqnarray}

It is obvious that the definition of the square root in (\ref{defsqrt}) is a special case of our general
definition for $q=1/2$.
The calculus of physical quantities outlined in section \ref{sec:CPQ}
can easily be extended from integer exponents to rational exponents
but we will not go into this in detail here.

\end{appendices}

\end{document}